\PassOptionsToPackage{table}{xcolor}

\documentclass[sigconf,authorversion,nonacm]{acmart}

\AtBeginDocument{%
  \providecommand\BibTeX{{%
    \normalfont B\kern-0.5em{\scshape i\kern-0.25em b}\kern-0.8em\TeX}}}

\PassOptionsToPackage{table}{xcolor}
\usepackage{wrapfig}
\usepackage{tikz}
\usepackage{fancyhdr}

\newcommand{\model}{IntFi~Model}
\newcommand{\fullmodel}{Interaction Fidelity Model}

\setcopyright{acmcopyright}
\copyrightyear{2024}
\acmYear{2024}
\acmDOI{XXXXXXX.XXXXXXX}

\acmConference[XXXX]{XXXX}{2024}{XXXX}
\acmPrice{15.00}
\acmISBN{978-1-4503-XXXX-X/18/06}

\begin{document}

    \emergencystretch=15pt

\title [The Interaction Fidelity Model]{The Interaction Fidelity Model: A Taxonomy to Distinguish the Aspects of Fidelity in Virtual Reality}

\author{Michael Bonfert}
\email{bonfert@uni-bremen.de}
\orcid{0000-0002-3605-6693}
\affiliation{%
  \institution{University of Bremen}
  \streetaddress{Bibliothekstraße 1}
  \country{Germany}
  \postcode{28359}
}
\author{Thomas Muender}
\email{thom@uni-bremen.de}
\orcid{0000-0002-0606-9258}
\affiliation{%
  \institution{University of Bremen}
  \streetaddress{Bibliothekstraße 1}
  \country{Germany}
  \postcode{28359}
}

\author{Ryan P. McMahan}
\email{rpm@ucf.edu}
\orcid{0000-0001-9357-9696}
\affiliation{%
  \institution{University of Central Florida}
  \streetaddress{4328 Scorpius Street}
  \country{USA}
  \postcode{32816-2362}
}

\author{Frank Steinicke}
\email{frank.steinicke@uni-hamburg.de}
\orcid{0000-0001-9879-7414}
\affiliation{%
  \institution{Universität Hamburg}
  \streetaddress{Vogt-Kölln-Str. 30}
  \country{Germany}
  \postcode{22527}
}

\author{Doug Bowman}
\email{dbowman@vt.edu}
\orcid{0000-0003-0491-5067}
\affiliation{%
  \institution{Virginia Tech}
  \country{USA}
  \postcode{24061}
}

\author{Rainer Malaka}
\email{malaka@uni-bremen.de}
\orcid{0000-0001-6463-4828}
\affiliation{%
  \institution{University of Bremen}
  \streetaddress{Bibliothekstraße 1}
  \country{Germany}
  \postcode{28359}
}

\author{Tanja Döring}
\email{tanja.doering@uni-bremen.de}
\orcid{0000-0001-8648-340X}
\affiliation{%
  \institution{University of Bremen}
  \streetaddress{Bibliothekstraße 1}
  \country{Germany}
  \postcode{28359}
}

\renewcommand{\shortauthors}{Bonfert, et al.}

\begin{abstract}
Fidelity describes how closely a replication resembles the original. It can be helpful to analyze how faithful interactions in virtual reality (VR) are to a reference interaction. In prior research, fidelity has been restricted to the simulation of reality---also called realism. Our definition includes other reference interactions, such as superpowers or fiction.
Interaction fidelity is a multilayered concept. Unfortunately, different aspects of fidelity have either not been distinguished in scientific discourse or referred to with inconsistent terminology. 
Therefore, we present the \fullmodel\ (\model). Based on the human-computer interaction loop, it systematically covers all stages of VR interactions. The conceptual model establishes a clear structure and precise definitions of eight distinct components.
It was reviewed through interviews with fourteen VR experts.
We provide guidelines, diverse examples, and educational material to universally apply the \model\ to any VR experience. We identify common patterns and propose foundational research opportunities.

\end{abstract}





\maketitle

\fancyfoot[C]{\thepage} 

\vspace{\baselineskip}
\textbf{This is a preprint that has not yet been published in a peer-reviewed journal. The manuscript is currently under review. \\Current version of figures, definitions, and supplemental material: v2.1}

\newpage
\section{Introduction}
Realism in virtual reality (VR) is pursued intensely in research and development~\cite{Rogers2022.suchwow, kapralos2014.SurgicalEducation, bowman2012.Naturalism,Gisbergen2019.RealismExpBehav, ragan2015.VisualComplexity, Grant2019.AnalysisOfRealism, witmersinger1998.PQ}. While the concept of \textit{realism}---how closely a simulation resembles reality---may initially seem straightforward, it is a complex, multi-faceted construct. 
We quickly assess something as realistic or unrealistic, be it a painting, the behavior of a movie character, a synthetic voice, or a virtual world. However, this intuitive judgment is insufficient for a comprehensive understanding of and reasoning why something is perceived as more or less realistic, especially in such a complex domain as VR, where countless factors might influence the outcome. To purposefully design virtual experiences that are convincingly realistic, it is essential to untangle the different aspects that impact the overall realism.

\begin{figure*}[t]
  \centering
  \includegraphics[width=1\textwidth]{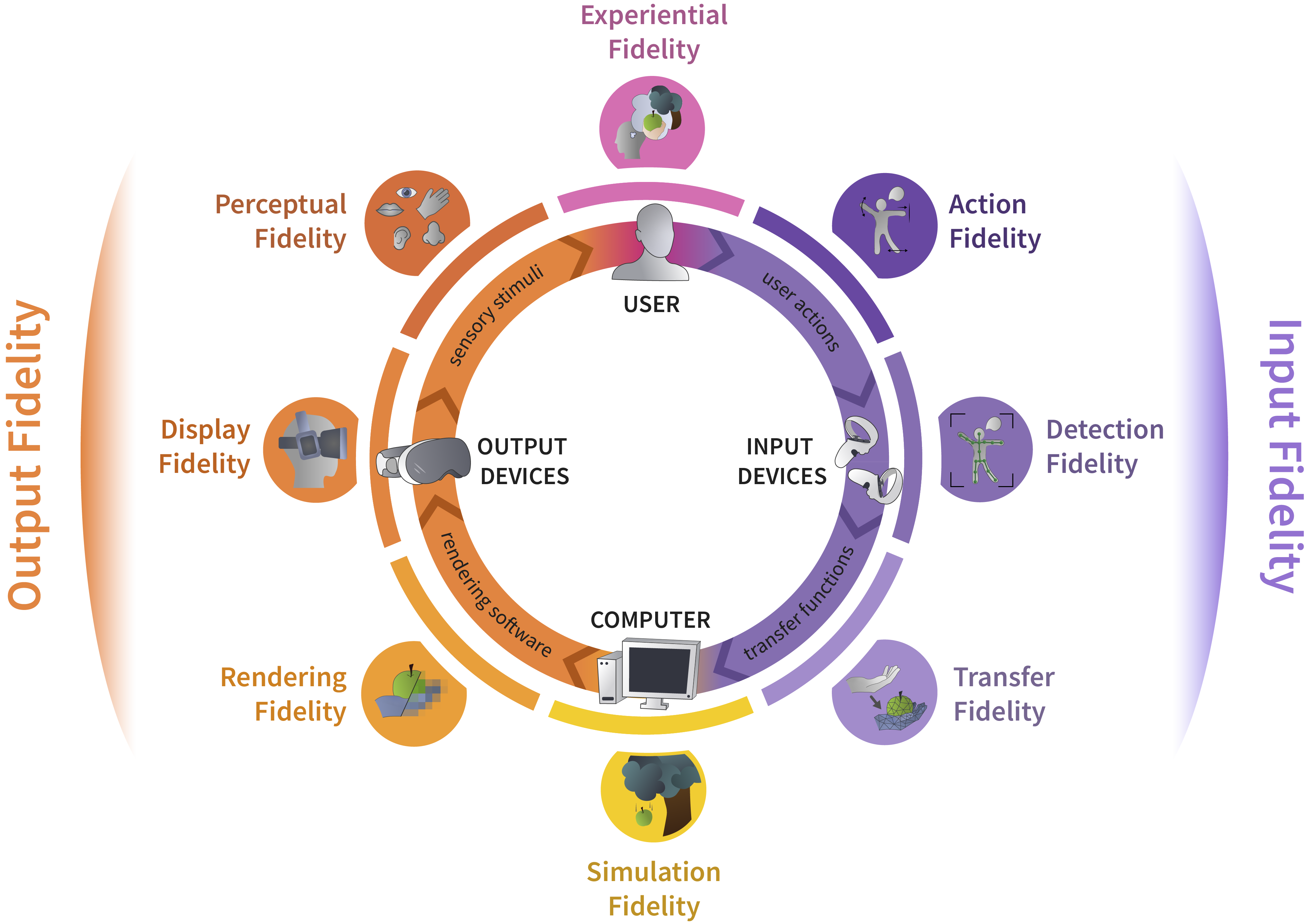}
  \caption{The {\fullmodel} differentiates between eight fidelity components that affect any VR interaction. The model is based on the Human--Computer Interaction Loop \cite{reenskaug_MVC_1979, norman1986.UserCenteredDesign} and extends a previous framework by \citet{mcmahan2016.UncannyInteration}.}
  \Description{An illustration of the \fullmodel. It is a circular arrangement of the following items in clockwise order starting at the top right: User actions and action fidelity, input devices and detection fidelity, transfer functions and transfer fidelity, the computer and simulation fidelity, rendering software and rendering fidelity, output devices and display fidelity, perception of sensory stimuli and perceptual fidelity, as well as the user and experiential fidelity. The components on the right side of the loop can be grouped as input fidelity and on the left side as output fidelity.}
  \label{fig:loop}
\end{figure*}

VR technology can create immersive experiences of being in and interacting with simulated realities. By interacting with the VR system, a user can perceive and affect the virtual environment (VE), while the system can sense and react to user input. As in the real world, users and their environments can mutually influence each other.
For many VR applications, realism is a decisive quality metric. The true-to-life resemblance is essential for skill training (e.g., surgery~\cite{chheang2019collaborative}), learning abilities (e.g., sports climbing~\cite{schulz2019role}, vocational education like public speaking~\cite{poeschl2017virtual}, music~\cite{serafin2017considerations}), entertainment (e.g., traveling the world~\cite{sarkady2021virtual}), therapy (e.g., fear of heights~\cite{freeman2018automated}), or use cases that would be expensive or impossible without VR (e.g., visiting Mars~\cite{Holt2023.VRforAstronauts}). In these scenarios, the success of the simulation depends on how closely the equivalent from reality can be reproduced. Even in fictional scenarios, certain aspects of the interaction might need to be grounded in reality (e.g., Euclidean geometry, spatial audio, swarm behavior, gravity, or the color space perceptible by humans). However, the concept of realism is limited to matching the real world and, therefore, cannot be applied to VR use cases simulating aspects impossible in reality. 

\begin{figure*}[ht]
  \centering
  \includegraphics[width=\linewidth]{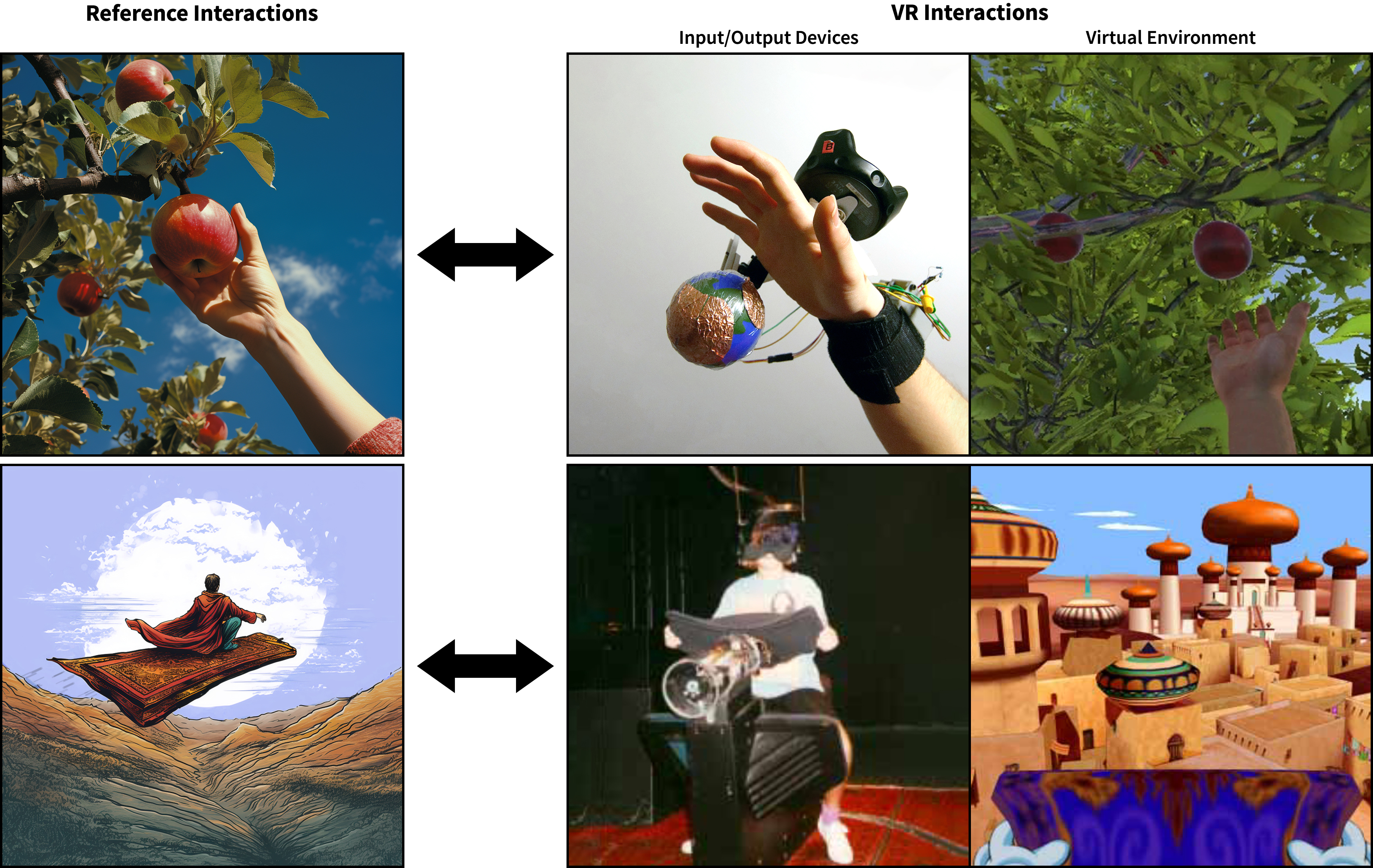}
  \caption{Two examples of reference interactions that are virtually replicated in VR simulations.  \textit{(top)} Here, an interaction from the real world is reproduced. Picking an apple from a tree is a complex activity to be simulated virtually as the user's actions, the provided feedback, and the physical simulation are closely coupled. This is a recent implementation with a haptic device by~\citet{deTinguy2020.weatavix} which dynamically moves the sphere proxy into the user's hand when grasping. \textit{(bottom)} Here, a fictional interaction is reproduced. The mythological \textit{magic carpet} from Middle Eastern literature can be realized as a VR locomotion technique in many different ways~\cite{medeiros2020.MagicCarpet}. This is an early implementation by \citet{pausch1996.Aladdin}. \textit{Copyright of the top-right images by~\cite{deTinguy2020.weatavix} and of the bottom-right images by~\cite{pausch1996.Aladdin} (modified by the authors). The images for the reference interactions have been generated with Midjourney.}}
  \Description{In the first row, the reference interaction of picking an apple is compared with a VR implementation. The first image is a photograph of a hand grasping and pulling down an apple. The second image shows the input/output devices where a device is mounted to the user's wrist with a Vive tracker and a sphere proxy approaching the palm. The third image of the virtual environment shows a detailed rendering of a virtual hand grasping for an apple.
  In the second row, the reference is a sketch of the concept of a flying rug with a person sitting on it flying over a mountain landscape. The depicted I/O devices are a bicycle-like construction on which a user with an HMD sits. The virtual environment shows a 3D-rendered scene resembling a Middle Eastern cityscape with dome-shaped towers, and in the foreground, there is a patterned flying carpet with two cartoony white hands holding the sides.}
  \label{Fig:correspondence}
\end{figure*}

\subsection{The Concept of Fidelity}
More generally speaking, the degree of how accurately an original is reproduced is called \textit{fidelity}~\cite{merriam-webster2024.FidelityDefinition}. 
Every VR simulation recreates some reference, be it a real-life situation, a training scenario, a fictional world, or a designer's imagination. From the Latin term \textit{fidēlis} for ``faithful,'' fidelity describes how faithful something is to its original. When simulating reality, this degree of correspondence is called realism. Thus, realism is a specific form of fidelity. Therefore, a VR application can have low realism yet high fidelity to a reference frame other than the real world. For example, the swords in \textit{Beat Saber} have high fidelity to lightsabers from \textit{Star Wars} but are unrealistic. Comparing simulations more generally to reference systems than only to the real world has been suggested by \citet{raser1969.SimulationSociety} in 1969, and we adopt this notion for more universal applicability. 
Figure~\ref{Fig:correspondence} compares reference interactions and their simulated VR interactions in two examples. On the top, a real-life activity is reproduced virtually: picking an apple from a tree. This has been investigated with haptic interfaces of different realism~\cite{kovacs2020.HapticPivot,deTinguy2020.weatavix}. On the bottom, the fantasy concept of flying with a magic carpet is adapted as a locomotion technique~\cite{pausch1996.Aladdin}. This metaphor affords a large design space of possible realizations with varying interaction fidelity in VR~\cite{medeiros2020.MagicCarpet}.  
The terms fidelity, realism, and naturalness are often used universally in scientific literature without detailing which aspect is referred to. If a specific aspect of fidelity is mentioned, a clear definition is often missing~\cite{Rogers2022.suchwow}. As a result, the terms and definitions within the VR literature have been inconsistent and contradictory, as illustrated in Section~\ref{sec:RW}. 

While the community's research efforts have led to useful definitions, models, and frameworks (see, e.g., Table~\ref{tab:signpost}), these mainly focus on dedicated aspects of fidelity, differ in their use of terms, and therefore do not provide a comprehensive understanding. 
This makes it harder to establish links between individual discoveries, generalize the results, and synthesize fundamental principles. Thus, this research aims to provide an umbrella framework that conflates existing findings on different aspects of fidelity into one comprehensive and consistent model.

\subsection{Introducing the \fullmodel}
Therefore, we present the \fullmodel\ (\model). This conceptual model distinguishes the different aspects of fidelity inherent in all VR interactions. 
The \model\ considers not only the system's fidelity but also the fidelity of interactions between the user and the system because of their reciprocal relationship. Beyond physical and functional simulation of a virtual environment in the form of bits and bytes, VR technology requires accounting for how the user's body affects the virtual world and how output devices can generate physical stimuli. This makes a holistic integration of all elements of embodied 3D interactions imperative.
Therefore, the \model\ is based on the human-computer interaction (HCI) loop~\cite{norman1986.UserCenteredDesign}, a well-established design principle that breaks down how a user and a system perceive and influence each other. 
For this, one aspect of fidelity is assigned to each of the eight stages of the HCI loop, as illustrated in Figure~\ref{fig:loop}. As a result, the \model\ consists of eight distinct fidelity components: (i) action, (ii) detection, (iii) transfer, (iv) simulation, (v) rendering, (vi) display, (vii) perceptual, and (viii) experiential fidelity. 

Building on prior work, the proposed model establishes a clear structure of the fidelity components with consistent terminology, precise definitions, detailed explanations, and illustrative examples in Section~\ref{sec:model}. 
This paper serves as a signpost by referring to more specialized frameworks and models detailing single components beyond the scope of this work. The \model\ can also help set a rigorous research agenda to advance purposeful measurement methods, determine factors contributing to fidelity, and understand the interdependence of the individual components. The \model\ can inform the VR community on how to focus its efforts to achieve a broad comprehension of realistic interactions. Beyond demonstrating how realistic a simulation is or how its fidelity differs from another, the model's theoretical foundation allows us to understand why~\cite{wobbrock2106.hciContributions}. Hence, theory-driven study designs facilitate more generalizable evaluation results and make linking them to other research easier.

The next section summarizes previous approaches, terms, and conceptualizations of fidelity and related constructs. We then present the \model\ in detail in Section~\ref{sec:model}. The subsequent section demonstrates the model's application with three example use cases. 
In Section~\ref{sec:validation}, we describe the validation process and report the findings from 14 semi-structured, one-hour interviews with VR experts from research and the industry. 
In Section~\ref{sec:bestpractices}, we share best practices for applying the model, explain how it can serve as different lenses through which VR interactions can be viewed, and caution against common application traps.
Section~\ref{sec:discussion} provides a discussion of typical fidelity patterns, practical and theoretical implications, and limitations of the proposed model. 
Lastly, we propose an abundant research agenda that the \model\ opens up and can inspire future work in our field in Section~\ref{sec:researchOps}.

\section{Related Work}
\label{sec:RW}
As soon as users enter a VE, they interact with the simulation. Within the scope of this paper, we consider interaction as a reciprocal exchange of a user and a computer system observing and reacting to each other through actions and states. The actions taken by users and the output generated by a computer depend on each other and together form the interaction. Therefore, interactions with a computer system cannot be attributed solely to humans or computers. The two must be considered together~\cite{hornbaek_interaction_2017}. The two-sided behavior happens simultaneously, continuously, inseparably, and inevitably, similar to a person affecting and being affected by the world around them in reality. 

\begin{figure}
  \centering
  \includegraphics[width=\columnwidth]{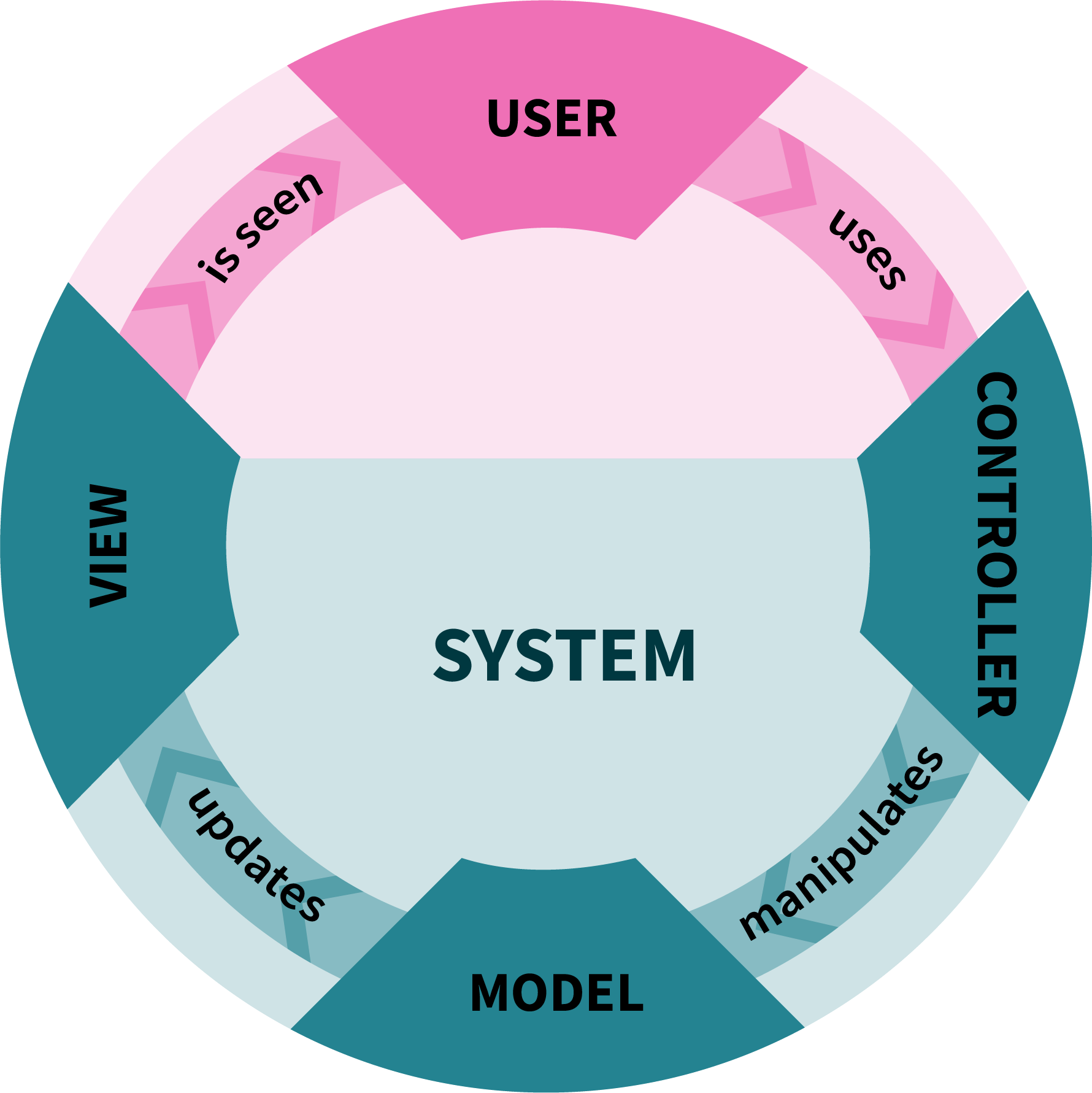}
  \caption{The Model-View-Controller pattern~\cite{reenskaug_MVC_1979} from 1979 on which the \model\ is conceptually based: A user uses a controller to manipulate the model, which updates the view and is then seen by the user. 
  }
  \Description{An illustration of the Model-View-Controller pattern. It is a circular arrangement that presents the following relations: A user uses a controller to manipulate the model, which updates the view and is then seen by the user. The items model, view, and controller represent the system. 
  }
  \label{Fig:MVC}
\end{figure}

Although this exchange simultaneously occurs in both directions, it can be helpful to think about the interaction as a circular sequence of steps for conceptually distinguishing them. 
A software design framework that describes this circular process is the model-view-controller (MVC) pattern introduced by \citet{reenskaug_MVC_1979} in 1979. This software architecture pattern became one of the most influential for describing and developing user interfaces. We illustrate the process in \autoref{Fig:MVC}. 
A design principle that describes a similar process with HCI-specific labels is what we call the \textit{human-computer interaction loop}. Forming the inner circle in Figure~\ref{fig:loop}, the HCI loop links the user, the input devices, the computer, and the output devices with transitions between these elements. This model can be traced back to a chapter by David Owen in ``User Centered System Design'' (p. 368) edited by Don Norman and Stephen Draper \cite{norman1986.UserCenteredDesign}. In the same book, Norman describes a similar process from a cognitive perspective of how users must overcome the Gulfs of Execution and Evaluation. He argues that the user continuously evaluates the current system state and plans actions to accomplish a specific goal---input and feedback.  Numerous frameworks and textbooks adapted the loop for different purposes and specializations, making it an established tool in HCI research~\cite{abowd1991.InteractionFramework, mcmahan2016.UncannyInteration,palanque2020.FaultsLoop, laviola2017.3DUI,williamson2009.BCIloop, mcmahan2018.SystemFidelity}.
Combined, the states and transitions of the HCI loop form the eight components of the proposed \model, and the two gulfs correspond to the grouping of input into and output from the system.

\subsection{Understanding Interaction Fidelity}
In the context of VR, the interaction with a system has the purpose for the user to experience and influence a simulated reality. As outlined in the introduction, the level of fidelity plays a vital role in many VR simulations. 
Several frameworks, models, and evaluations have investigated various aspects of interaction fidelity. Please note that we refer to the original terms from cited works instead of the \model's terminology in this subsection. As a result, they might appear confusing and contradictory.

Most prominently, \citet{mcmahan2011exploring} proposed the \textit{Framework for Interaction Fidelity Analysis} (FIFA) that was released in an updated version in 2016~\cite{mcmahan2016.UncannyInteration}. The revised FIFA considers the three categories \textit{biomechanical symmetry} describing the reproduction of body movements from the real world, \textit{input veracity}, which considers the exactness of input devices capturing movements, and \textit{control symmetry}, which covers the exactness of control in the virtual world compared to the real world. Each category comprises further detailed components. The framework is designed to compare the user's motions during virtual activities that involve body movements, such as techniques for locomotion or object manipulation, to their counterpart in reality. Results from their user studies suggest an uncanny valley of VR interactions. They found good user performance with low- and high-fidelity systems but a drop in performance with medium-fidelity systems. Similar findings have been presented in further studies~\cite{bhargava2018.IntFiContinuum, nabiyouni2015.LocomotionTechniques}.
In other investigations, \citet{bowman2012.Naturalism} also found that high-fidelity interactions can enhance performance and the overall user experience, but medium levels of fidelity can be unfamiliar and detrimental to performance. 

The FIFA framework, however, only considers the fidelity of user actions and, therefore, only the input side of the two-way interaction. The system output is neglected in this framework even though it is an inseparable part of the interaction and can heavily influence the realism of a system. \citet{mcmahan_DisplayIntFidelity_2012} acknowledge the missing output component by also analyzing the display fidelity and finding a similar negative effect for medium display fidelity as for their interaction fidelity. \citet{nilsson_DecreasedFidelity_2017} build on this finding and argue that when some fidelity components are limited, maximizing the fidelity of other components may be detrimental to the perceived realism. Therefore, decreased fidelity might positively influence perceived realism in some instances. In addition, \citet{abtahi2022beyond} show that even for interactions that go beyond reality, some aspects of the interaction should still be grounded in the real world to avoid sensory conflicts in the user. 

For evaluating the fidelity of a VR simulation, \citet{stoffregen2003nature} consider the action fidelity as the relationship between performance in the simulator and performance in the simulated system, the system's output in the form of optic, acoustic, mechanical, and inertial arrays, as well as the experiential fidelity in the form of perceived presence~\cite{skarbez2017.Presence}. 
With a focus on more practical aspects of current VR systems, \citet{al2022framework} present a framework for evaluating fidelity concerning four interrelated elements: digital sensory system fidelity, interaction system fidelity, simulation system fidelity, and integration among these aspects to produce high-fidelity virtual experiences. They identify various factors for evaluating VR hardware regarding visual, auditory, and haptic feedback, the tracking system, and graphic quality. Conversely to the FIFA framework, this framework focuses on the output side of interactions and neglects the user actions as part of the interaction with the VR system. The fidelity of haptic feedback can also be assessed in more detail with the Haptic Fidelity Framework by \citet{hapticfidelityframework}, providing detailed factors to analyze and quantify aspects of sensing, hardware, and software.

In the context of gaming, \citet{rogers2019.InteractionFidelity} evaluated interaction fidelity for object manipulation and whole-body movements and found that high fidelity is preferred for object manipulation. Still, moderate fidelity can suffice for whole-body movements as there is a trade-off between fidelity, usability, and social factors. Further, \citet{Rogers2022.suchwow} provide an in-depth analysis of realism in digital games, including a focus on VR. The authors present a two-part framework of realism dimensions consisting of a hierarchical taxonomy of realism dimensions and the mapping of realism dimensions within Adams' game model~\cite{adams2014.GameDesign}. 
\citet{alexander2005gamingTraining} investigated the effect of fidelity on the transfer of knowledge from games and simulations to the real world. They argue that the fidelity of a simulation is a significant factor in enabling skill transfer and define three categories of fidelity: Physical fidelity is the degree to which the simulation looks, sounds, and feels like the real world; functional fidelity is the degree to which the simulation acts like the real world; and psychological fidelity is the degree to which the simulation replicates the psychological factors (e.g., stress, fear) experienced in the real world. 

The \textit{Reality-Based Interaction} framework by \citet{jacob2008reality} provides four themes to enable high-fidelity interactions on a more general level with interfaces such as touchscreens, tangibles, and VR. Interaction designers should consider naïve physics, body awareness and skills, environment awareness and skills, and social awareness and skills. The work outlines trade-offs between realism and expressiveness, efficiency, versatility, ergonomics, accessibility, and practicality. 
In contrast to most other frameworks that cover input and output components of the interaction, \citet{experientialfidelity} focus on experiential fidelity, enhancing the realism of the user experience by guiding the user's frame of mind in a way that their expectations, attitude, and attention are aligned with the VR experience.

\subsection{Inconsistent Fidelity Terminology}
To this point, we have adhered to the terminology originally used in the mentioned works. The literature established a patchwork of different but similar terms based on different interpretations and assumptions. This is why the wording of the above explanations might sound inconsistent and contradictory. It demonstrates how critical uniform designations are for research communication. 

The terms fidelity, realism, and naturalness were often used synonymously in previous literature. Researchers often investigated only a specific part of interaction fidelity but referred to it universally as (interaction) fidelity. 
For example, the term \textit{interaction fidelity} has been used to refer to visual render quality~\cite{Mania2006}, camera views and gravity~\cite{bhargava2018.IntFiContinuum}, or dialogue capabilities~\cite{Carnell2022}. Some publications refer only to the user's system input with it \cite{mcmahan2016.UncannyInteration, bowman2012.Naturalism}. This neglects half of the two-way interaction between the user and the system, which can only be considered in its reciprocal dependence, as outlined at the beginning of this section. Also, the literature generally refers to other individual aspects of the interaction as \textit{fidelity}. 
For example, some fidelity conceptions focus on the simulated virtual environment, such as in game research~\cite{lukosch2019.FidelityGaming, alexander2005gamingTraining}, or are reduced to the simulation's physical and functional dimensions~\cite{hays1988.SimulationFidelityTraining, harteveld2011.TriadicGameDesign}, while it is crucial for VR and 3D interfaces also to consider the means of input and output as well as the user's role. A recent framework classified the fidelity of mixed-reality prototyping~\cite{cox2022.PrototypingFidelity}. 
Furthermore, outside computer science, fidelity has been narrowly defined within the fields' contexts, such as in health and psychology regarding realistic psycho-behavioural and affective responses~\cite{harris2020.ValidationFramework,arthur2023.FidelityInHealth}. 
These examples illustrate how divided the VR community has been about the term's understanding and usage.

In a systematic review of the concepts of realism and fidelity for digital games, \citet{Rogers2022.suchwow} found a ``substantial potential for confusion given the overlapping and contradictory use of realism types.'' The authors report that the type of realism is often not even further defined but remains vague in the literature. The rigorous analysis covers VR research as part of gaming but excludes the realism of other VR interactions and the fidelity compared to other reference frames.
Nevertheless, the survey outlines the vast range of terms used to describe aspects of realism and fidelity.
This emphasizes the urgent need for a theoretical basis of consistent terminology. Plenty of research contributes to the understanding of the multidimensional concept of fidelity. Still, it lacks an umbrella model into which the individual elements can be integrated to understand the bigger picture.
We will consequently use the \model's terminology for the remainder of this paper.

\begin{figure*}[ht]
  \centering
  \includegraphics[width=0.9\textwidth]{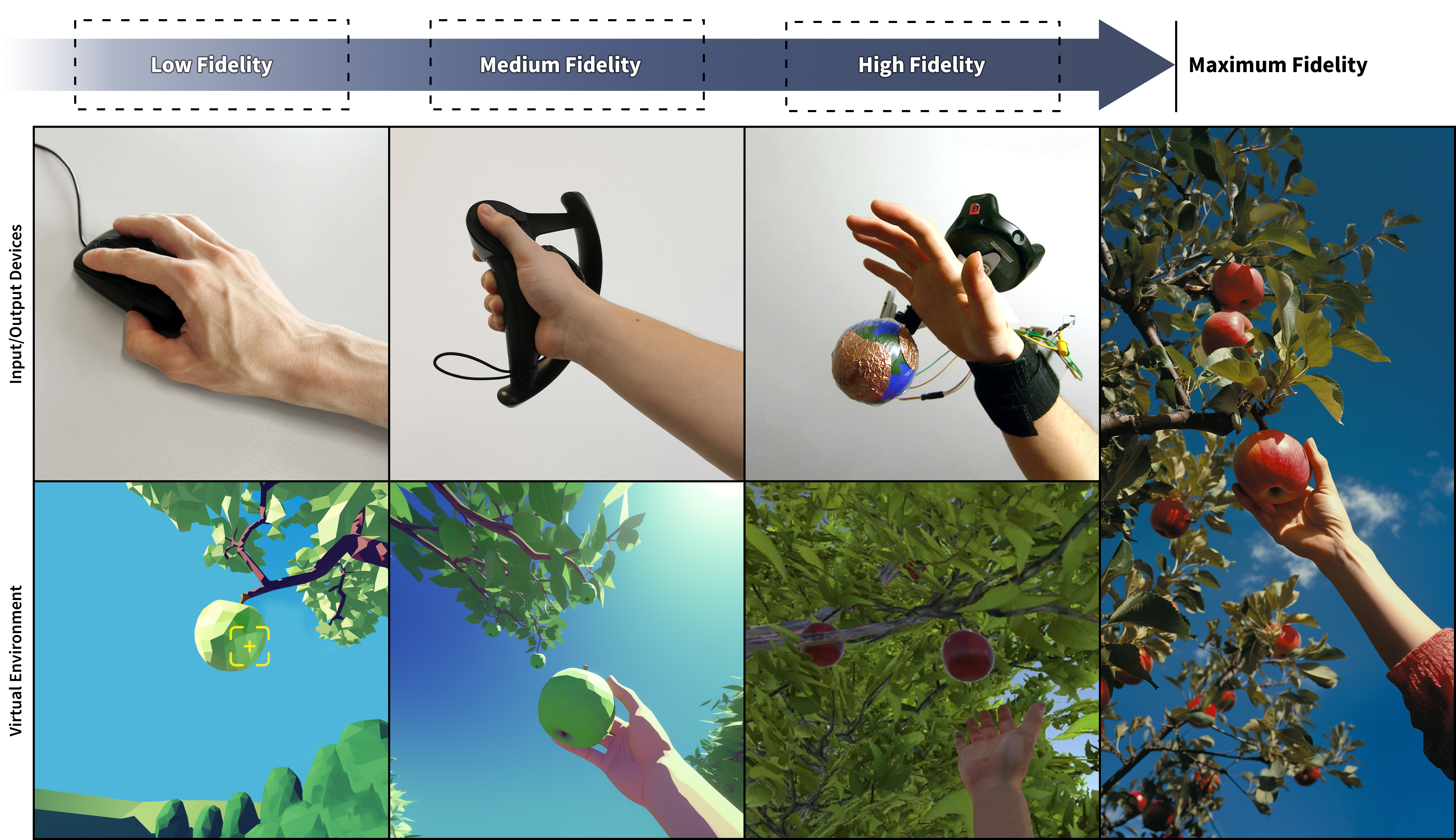}
  \caption{The fidelity spectrum with approximate ranges from low, medium, and high to maximum fidelity with an example use case: implementations with different fidelity levels of somebody picking an apple from a tree. The reference interaction from the real world on the right side is defined as maximum fidelity.
   \textit{Copyright of the ``High Fidelity'' images by ~\citet{deTinguy2020.weatavix}. The other images have been photographed or generated with Midjourney by the authors.}}
  \Description{An arrow is labeled with the ranges low, medium, and high fidelity with maximum fidelity at the end. Aligned with these ranges, example implementations of a person picking an apple from a tree with the respective fidelity range are presented. In the low-fidelity example, a user selects the apple with a computer mouse, indicated with a crosshair in the low-poly virtual environment. 
  In the medium-fidelity example, the user grasps the apple with a hand-held Valve Index controller, represented by a low-poly hand avatar in the virtual environment.
  The high-fidelity example shows the system Weatavix by \citet{deTinguy2020.weatavix}. Here, the user grasps a dynamic, wrist-mounted sphere proxy providing passive haptic feedback, represented with detailed visuals in the virtual environment.
  On the right side, maximum fidelity is illustrated with the reference interaction from reality: a person picking an apple from a tree in the real world.}
  \label{fig:spectrum}
\end{figure*}

\section{Model of Interaction Fidelity}
\label{sec:model}
The conceptual model presented here distinguishes various aspects of the fidelity of interactions in VR. It covers the entire process of a user interacting with a VR system, from user input over system processing to output from the system experienced by the user. 
Instead of assessing the contribution of each device or system component to the fidelity of the interaction, we propose distinguishing between the stages of the interaction to systematically evaluate how true it is to the original.
The \model\ is based on the HCI loop~\cite{norman1986.UserCenteredDesign}, which originates from the model-view-controller pattern described in Section~\ref{sec:RW}. Following the structure of the loop, the model assigns one aspect of fidelity (for example, display fidelity) to one stage of the loop (in this example, output devices), as illustrated in Figure~\ref{fig:loop}.
The loop offers simplicity, yet all fidelity aspects of any conceivable interaction are integrated. Therefore, it is a sound foundation for the intuitive differentiation of factors that define the fidelity of any VR interaction with the user in mind.

Based on the Merriam-Webster Dictionary~\cite{merriam-webster2024.FidelityDefinition}, \citet{mcmahan2011exploring}, \citet{alexander2005gamingTraining}, and \citet{raser1969.SimulationSociety}, we consider \textit{interaction fidelity} as the degree of exactness with which reference interactions are reproduced. Thus, it describes how closely a user's interactions with a VR system resemble the interactions from a reference system. This reference system can be the real world, in which case we refer to realism, but we can also choose any other reference interaction, such as fictional worlds (e.g., Star Wars), hyper-realistic interaction techniques (e.g., the Go-Go technique~\cite{poupyrev1996.GoGo}), a previous VR system, a planned system iteration, or a replicated study. It is important to clearly define the chosen reference interaction for a meaningful and unambiguous fidelity assessment. If the reference is changed to make another comparison, the assessed fidelity will also change.

We can describe the level of fidelity on a spectrum covering low, medium, and high to maximum fidelity as illustrated in Figure~\ref{fig:spectrum}. With maximum fidelity, there is theoretically a perfect correspondence to the original, even if it might be technologically impossible to achieve. Between perfect and no correspondence, there is a continuum~\cite{bowman2007.EnoughImmersion,bhargava2018.IntFiContinuum, lukosch2019.FidelityGaming, cox2022.PrototypingFidelity} without clear-cut ``low'', ``medium'', or ``high'' states. This wording demonstrates a relative difference or approximate range on the continuum.

It is crucial to keep in mind that fidelity is an objective concept simply describing the degree of correspondence without judgment. 
Higher interaction fidelity is not necessarily better, more desirable, more effective, or more immersive but merely implies a closer match to the reference. Although higher fidelity can have benefits for other metrics or goals, it has also been shown how lower-fidelity and hyper-natural interactions can be beneficial~\cite{McMahan2010,feinstein2001.FidelityVerifiabilityValidity,hays1992.FlightSimEffectiveness,dewitz2023.MagicInteractionsFramework,nabiyouni2015.LocomotionTechniques, mcveigh-schultz2021.WeirdSocialSuperpowers}.
On the other hand, aspects of fidelity often determine the success of a simulation. For instance, in motor skill learning, the faithfulness of the user's movements is crucial. Likewise, authentic scenic details are the key aspect of a travel simulation, accurate haptic feedback during surgical training, and plausible situations for phobia therapy. As Section~\ref{sec:discussion} outlines, objective system fidelity does not necessarily correlate with perceived realism. For effective and economical planning, interaction designers and VR developers must reflect on what kind of fidelity is important for the use case. 

All aspects of fidelity can be assessed objectively and subjectively depending on the point of view. When applying standardized metrics for reproducible, indisputable measures to describe fidelity, we can objectively determine and verify the exactness of the interaction's match with the reference interaction. For example, we can impartially compare two screens regarding their technical specifications, such as pixel density. When relying on personal impressions from interviews or questionnaires, we can subjectively assess fidelity. For example, we can ask a user in a questionnaire how closely the hand movements in a VR juggling training experience match juggling in reality. 
Some fidelity aspects can be assessed objectively in a more meaningful way, such as system specifications (e.g., screen resolution) or historical facts (e.g., the 1988 ACM Turing Award recipient) with verifiable ground truth. For other aspects, on the other hand, it might make sense to assess them subjectively, such as perception or experience. Currently, we do not have the means to determine every aspect objectively. This might be technically possible in the distant future, even for experiential fidelity, with sufficiently sophisticated brain--computer interfaces.

One could argue that users' perceptions and experiences are inherently subjective because they vary between individuals. However, while they are different between users, we can assess perceptual and experiential fidelity individually: How would the same person probably perceive and experience the reference interaction? Because systems are usually not tailored to individuals, typical or average users from a target group can be considered for better generalizability. For this pragmatic reason, we advocate for a population-centric assessment through user-centric research. 
Some fidelity aspects can be assessed independently (e.g., rendering fidelity), but especially for the user-related aspects (i.e., perceptual, experiential, and action fidelity), the target users' abilities and characteristics must be considered to provide accessible systems acknowledging the diversity of users. For example, people with color vision deficiency perceive the same visual output differently, which might affect how closely it resembles their real-life perception. Similarly, target populations can experience a system's fidelity differently, for example, depending on their expertise and how competent they feel in a virtual experience compared to their real-world competence. For example, experienced soccer players feel more restricted than novices in virtual kicking with medium simulation fidelity~\cite{bonfert2022.kicking}.
In discourse, we must keep in mind that we all perceive the world subjectively and create a mental model of how it works~\cite{norman1989.EverydayThings}. While people can have different perspectives on theoretical ground truth and be challenged in their view, we need to agree in discussions on an explicit reference interaction that is supposed to be simulated and target user groups for meaningfully applying the term fidelity.

\subsection{Development of the Model}
We have devised the idea for the \model\ when gathering different aspects of fidelity from the literature and testing possible classifications to bring structure to the concept. When cross-referencing the first approaches with related work, we realized that any dimension fits neatly into the HCI loop. In an iterative process, we refined the labels and definitions of the components from discussions among the authors, with research peers, in teaching practice, and at conferences. 
We conducted semi-structured expert interviews with 14 VR researchers and practitioners to improve and validate the model. We present the method and results in detail in Section~\ref{sec:validation}.

\subsection{Structure of the Model}
The model consists of the user and the VR system, which includes input devices, the computer as the processing unit with data and models, and output devices. 
Between these components, \autoref{fig:loop} uses arrows indicating a translation from software to hardware (such as the rendering from the simulation to the output devices) or, vice versa, from physical to intangible information (such as from the system output to the user's mind through perception). 
Every aspect of interaction fidelity corresponds to one stage of the HCI loop and is, thus, represented by one component in the model, as visualized in \autoref{fig:loop}. For example, the fidelity with which the system detects the user actions corresponds to the \textit{input devices} in the HCI loop, which is the \textit{controller} in the MVC paradigm and is linked to the \textit{detection fidelity} in this model. The single components of the model are further detailed in this section.

\renewcommand{\arraystretch}{1.5}
\begin{table*}
\centering
\caption{The definitions of the fidelity components and the categorizing sets of components.}
\label{tab:definitions}
\begin{tabular}{| m{.06\textwidth} m{.17\textwidth}  m{0.47\textwidth} | m{.22\textwidth}|}
\hline
\cellcolor[HTML]{B0B0B0} & {\cellcolor[HTML]{B0B0B0}\textbf{Aspect}} & \cellcolor[HTML]{B0B0B0} \textbf{is defined as the degree of exactness with which…} & \cellcolor[HTML]{B0B0B0} {\textbf{Depends on}} \\ \hline

\cellcolor[HTML]{7246a6} \includegraphics[height=30pt]{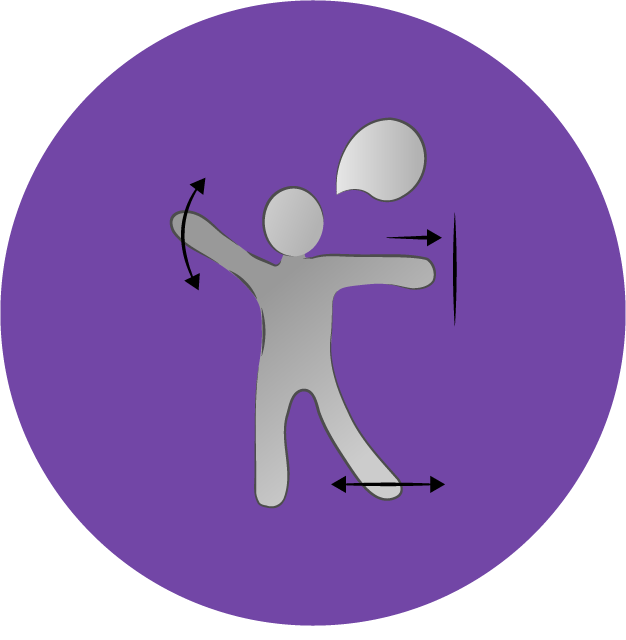} & \cellcolor[HTML]{7246a6} \textcolor{white}{\textbf{Action Fidelity}} & user actions resemble those of the reference interaction.  & User \\
\cellcolor[HTML]{8e6eb5} \includegraphics[height=30pt]{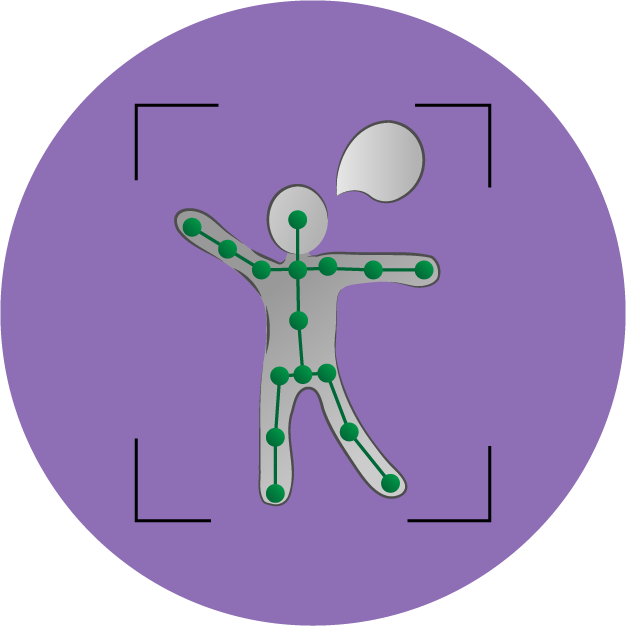} & \cellcolor[HTML]{8e6eb5} \textbf{Detection Fidelity} & input devices detect the user's actions. & System \\
\cellcolor[HTML]{ac8ecf} \includegraphics[height=30pt]{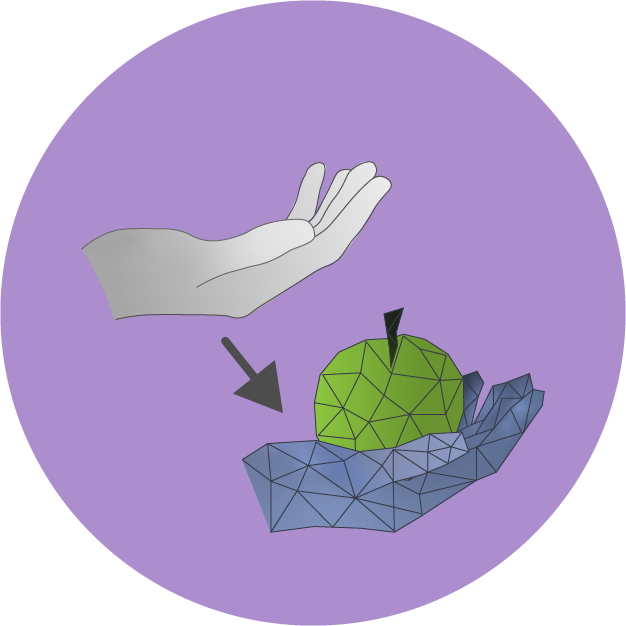} & \cellcolor[HTML]{ac8ecf} \textbf{Transfer Fidelity} & virtual actions, derived from the input measurements, resemble the user's actions of the reference interaction. & System \\ 
\cellcolor[HTML]{ffcf0c} \includegraphics[height=30pt]{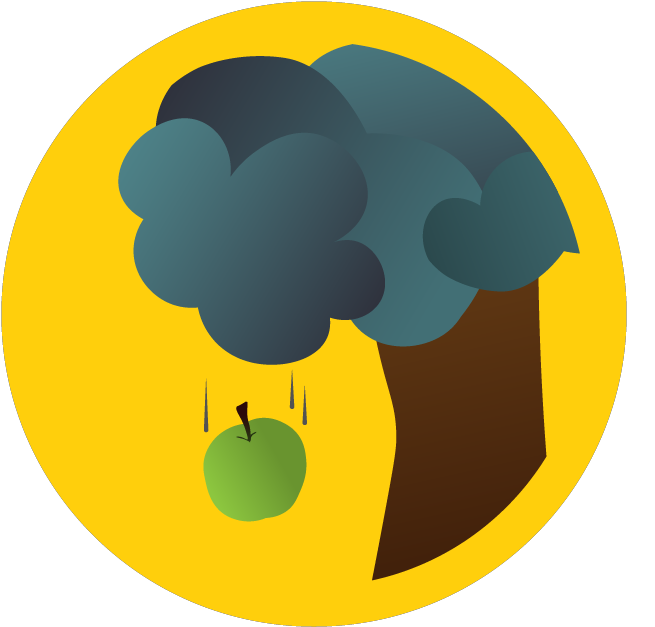} & \cellcolor[HTML]{ffcf0c} \textbf{Simulation Fidelity} & a virtual environment resembles the characteristics of the reference interaction's world and adequately reacts to the user's actions. & System \\ 
\cellcolor[HTML]{ffa12d} \includegraphics[height=30pt]{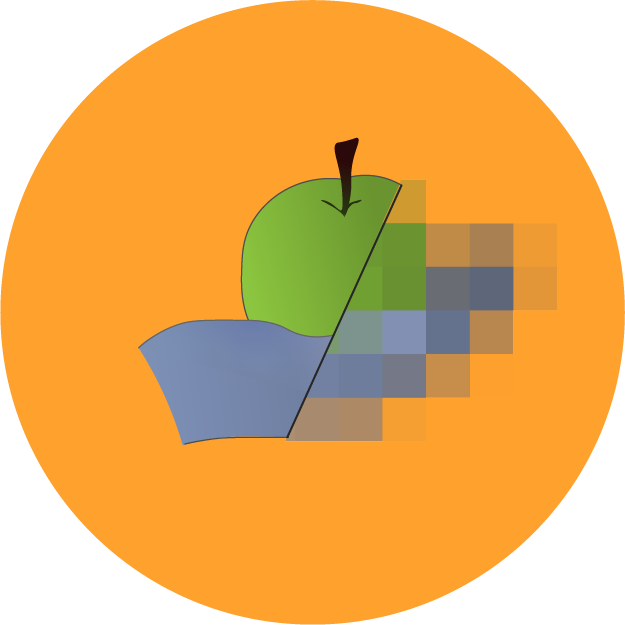} & \cellcolor[HTML]{ffa12d} \textbf{Rendering Fidelity} & the output content generated by the computer resembles what would be presented to the user in the reference interaction. & System \\
\cellcolor[HTML]{fb8739} \includegraphics[height=30pt]{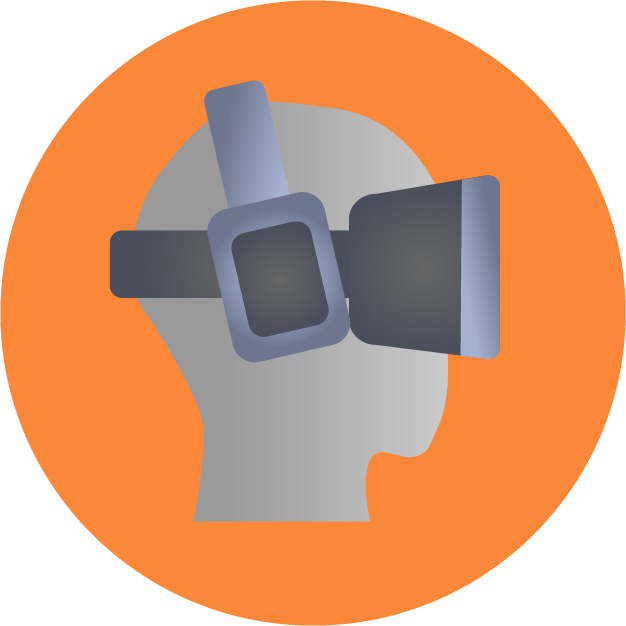} & \cellcolor[HTML]{fb8739} \textbf{Display Fidelity} & the output devices reproduce the physical stimuli presented to the user in the reference interaction. & System \\
\cellcolor[HTML]{ed7038} \includegraphics[height=30pt]{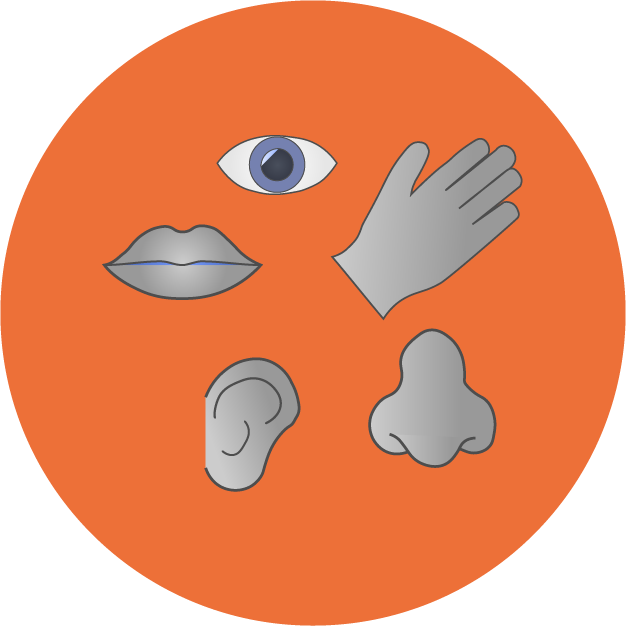} & \cellcolor[HTML]{ed7038} \textbf{Perceptual Fidelity} & the user's perception of the physical stimuli created by the system resembles how the user would perceive the reference interaction. & User \\ 
\cellcolor[HTML]{ef70b6} \includegraphics[height=30pt]{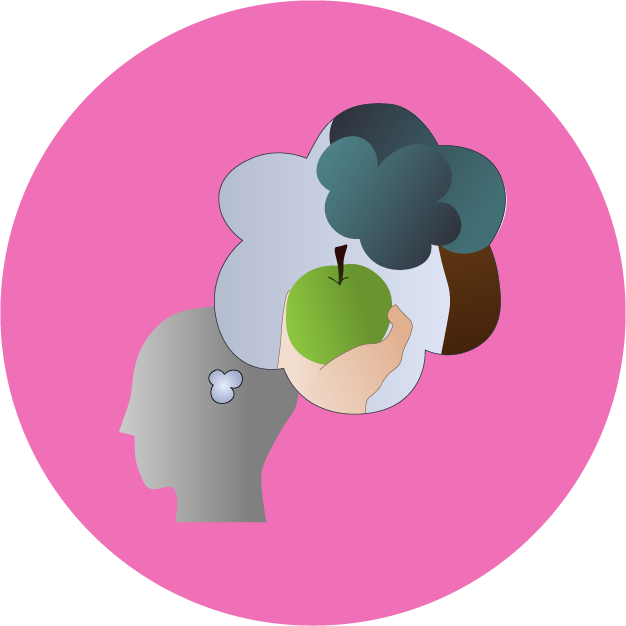} & \cellcolor[HTML]{ef70b6} \textbf{Experiential Fidelity} & the user's experience of the simulated interaction resembles how the user would experience the reference interaction. & User \\ 
\hline \hline

\cellcolor[HTML]{B0B0B0} & {\cellcolor[HTML]{B0B0B0}\textbf{Set of aspects}} & \cellcolor[HTML]{B0B0B0} \textbf{is defined as the degree of exactness with which…} & \cellcolor[HTML]{B0B0B0} {\textbf{Includes}} \\  \hline
\cellcolor[HTML]{000000} \includegraphics[height=30pt]{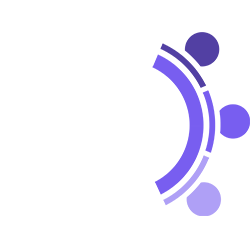} & \cellcolor[HTML]{9e72d4} \textbf{Input Fidelity} & the virtual actions generated from the user's input resemble the user actions of the reference interaction. & Action, Detection, Transfer \\
\cellcolor[HTML]{000000} \includegraphics[height=30pt]{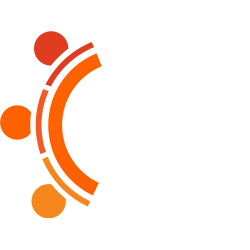} & \cellcolor[HTML]{fb8739} \textbf{Output Fidelity} & the system output is generated and perceived as the user would perceive it in the reference interaction. & Rendering, Display, Perceptual \\
\cellcolor[HTML]{000000} \includegraphics[height=30pt]{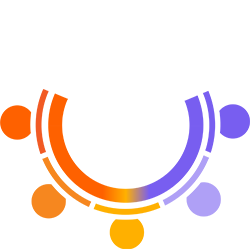} & \cellcolor[HTML]{52a8b3} \textbf{System Fidelity} & the system reproduces the world of the reference interaction reacting to the user. & Detection, Transfer, Simulation, Rendering, Display \\ \hline
\cellcolor[HTML]{000000} \includegraphics[height=30pt]{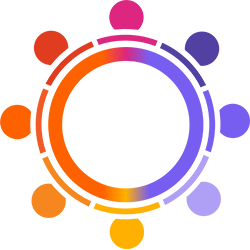} & \cellcolor[HTML]{000000} \quad \newline \textcolor{white}{\textbf{Interaction Fidelity}} \newline &  \textbf{reference interactions are reproduced.}  & \textbf{All aspects}  \\ \hline
\end{tabular}
\end{table*}

We can consider the model vertically and distinguish aspects of \textit{input fidelity} (right side) and \textit{output fidelity} (left side). We can also consider the model horizontally and distinguish aspects of fidelity that concern the \textit{user} with their perceptions, experiences, and actions (upper part) and those concerning \textit{system fidelity} with the detection, transfer, simulation, rendering, and displays (lower part).
Input fidelity in this model is close to what \citet{mcmahan2011exploring} described as interaction fidelity in the Framework for Interaction Fidelity Analysis (FIFA) and comprises similar components: \textit{action}, \textit{detection}, and \textit{transfer fidelity}. All further components were not considered in FIFA.

Aspects of fidelity that determine the characteristics of the VE and react to the user input are included in the component \textit{simulation fidelity}. Proceeding in the loop, the group of \textit{output fidelity} components comprises \textit{rendering fidelity}, \textit{display fidelity}, and \textit{sensory fidelity}. 
Finally, all these aspects combined determine \textit{experiential fidelity}, the impression created in the user's mind. 
In the \model, we focus on the endpoints of the single components instead of elaborating on the technical processes behind each component. For example, detection fidelity is determined by the final output of the input device's API, not by the sensor's firmware or signal processing. 
The single components can be broken down further as needed (e.g., interaction fidelity → simulation fidelity → presentational fidelity → 3D model → skin texture → height map → resolution). In the scope of this paper, we will only detail conceivable subcomponents through examples, not comprehensively, except for \textit{simulation fidelity}. 
As a whole, all aspects of the model define overall interaction fidelity.
In the following, the components are explained in detail following this structure: We define the fidelity aspect, distinguish it from the subsequent aspect, detail its characteristics, state its requirements for maximum fidelity, illustrate how different levels of fidelity could be designed, and refer to specialized frameworks or similar definitions.

\begin{wrapfigure}{l}{0.12\columnwidth}
    \includegraphics[width=0.169\columnwidth]{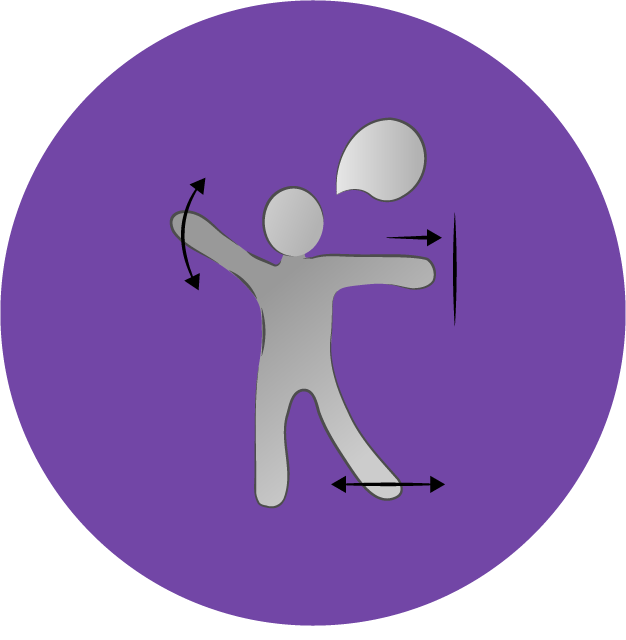}
\end{wrapfigure}
\subsubsection{\textbf{Action Fidelity}} 
Action fidelity is the degree of exactness with which user actions resemble those of the reference interaction. In the context of this model, we consider actions as any active behavior or passive state of the user, including sheer existence. 
In contrast to the next component, \textit{detection fidelity}, it is irrelevant for action fidelity whether the system captures the user actions. 

This component comprises all behavior and states of the user, including body movements, such as grasping or walking, and any other modality and activity, such as speaking, eye gaze, facial expressions, or brain activities. It is crucial to consider the full range of user actions. Not only intentional actions (e.g., gestures) are relevant for action fidelity but also subconscious (e.g., blinking), uncontrolled (e.g., blushing, swelling), involuntary (e.g., tremor), passive (e.g., static poses), and unaware actions (e.g., body temperature). 

In the example of grasping an object, the user of a low-fidelity solution would point at the object with a 3-degrees-of-freedom (DoF) controller and hold it by pressing a button. With high action fidelity, the user would reach out the hand to the object's position, enclose it with the fingers according to its shape and size, and exert force with the arm in proportion to its weight.
To virtually reproduce interactions with maximum fidelity, the user must behave exactly as in the reference interaction. 

This component is close to the \textit{biomechanical symmetry} dimension of the revised FIFA by ~\citet{mcmahan2016.UncannyInteration}. In their definition, only the movements of body parts are considered user actions. They distinguish between anthropomorphic, kinematic, and kinetic symmetry of body movements. 
We include additional forms of user actions in this model beyond active bodily motion.

\begin{wrapfigure}{l}{0.12\columnwidth}
    \includegraphics[width=0.169\columnwidth]{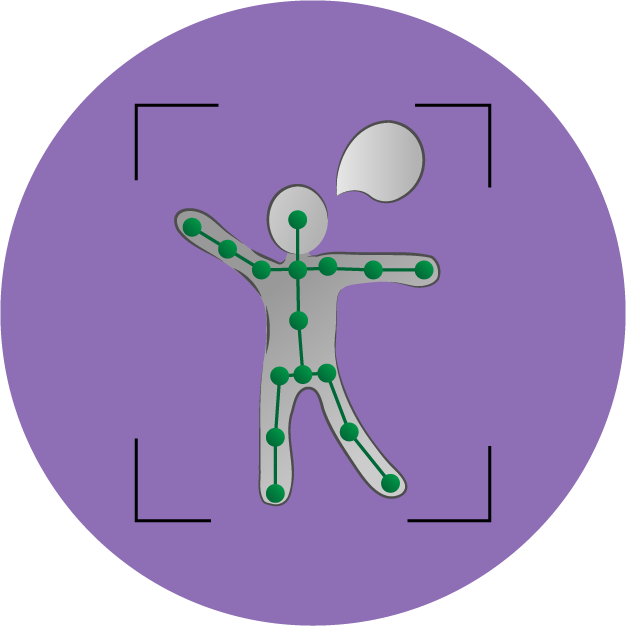}
\end{wrapfigure}
\subsubsection{\textbf{Detection Fidelity}} 
Detection fidelity is the degree of exactness with which input devices detect the user's actions.
In contrast to the next component, \textit{transfer fidelity}, it is irrelevant for detection fidelity how the system interprets or classifies the captured sensor data. While detection fidelity is about converting physical actions into measurements of those actions, transfer fidelity is about converting those measurements into meaningful virtual actions.

This component involves the capability of sensors to capture relevant signals. For high detection fidelity, the measurements are accurate, noiseless, reliable, and immediate. The kind of signal and the appropriate input device depend on the respective user actions and concerns the same range outlined under action fidelity, from optical tracking of body movements to microphones for speech detection to electromyography (EMG) for facial expressions. The sensor range can cover all possible parameters, such as thermosensors for measuring body temperature. The system does not need to detect actions or body signals that would not affect the reference interaction, such as the temperature of the feet when knitting with one's hands.

For example, a low-fidelity solution in a horror game would detect the user jumping in terror and gasping. In contrast, in a high-fidelity solution, the system would also detect the user's eyes snapping open, their muscles contracting, and the intensity of their sweating. 
To virtually reproduce interactions with maximum fidelity, the input devices must detect every relevant user action and state precisely. 

This component is close to the \textit{input veracity} dimension of the revised FIFA~\cite{mcmahan2016.UncannyInteration}. The authors distinguish between the measurement's accuracy, precision, and latency as the delay before sensory feedback. 
We consider the immediacy of the detection as a decisive factor but independent of system feedback.

\begin{wrapfigure}{l}{0.12\columnwidth}
    \includegraphics[width=0.17\columnwidth]{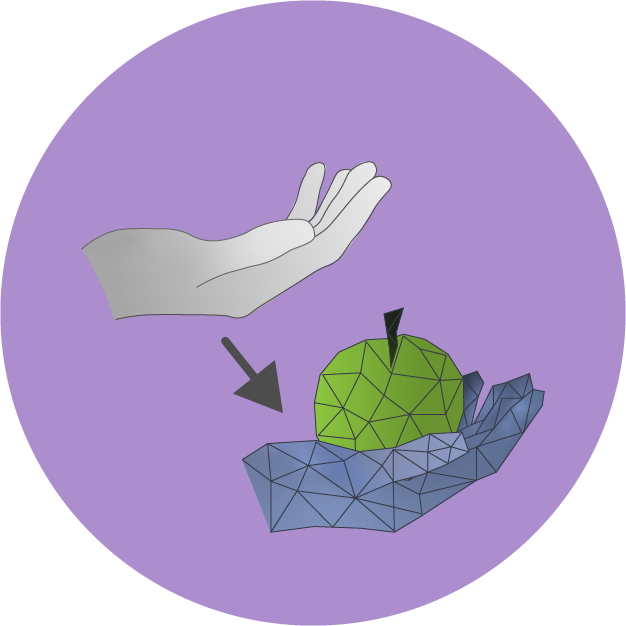}
\end{wrapfigure}
\subsubsection{\textbf{Transfer Fidelity}} 
Transfer fidelity is the degree of exactness with which virtual actions, derived from the input measurements, resemble the user's actions of the reference interaction.
In contrast to the next component, \textit{simulation fidelity}, it is irrelevant for transfer fidelity how the virtual actions affect the simulation. 

This component refers to how the system considers the sensor readings, processes and transforms them, and interprets them in the context of the simulation to generate virtual actions mapped from the user's real actions in a meaningful way. 
The virtual actions can deliberately differ from the real actions, e.g., by modifying the control/display ratio. Therefore, this component is determined by the correspondence of the virtual actions with the reference interaction to be simulated, not with the actual user actions, if these differ. For example, in redirected walking~\cite{razzaque2001.RedirectedWalking}, we compare the virtual path to the intended straight path, not the circular path that the user physically takes.
In case of incomplete, distorted, or simplified input data, processing can make up for input deficiencies to allow finding a probable interpretation.

As an example of low transfer fidelity, when processing noisy capacitive sensor data of a hand-held controller for approximating the hand pose, a low-fidelity solution would show jittery and anatomically absurd finger movements. In contrast, a high-fidelity solution would result in smooth and plausible finger movements that match the real hand pose of the user, e.g., using inverse kinematics. 
To virtually reproduce interactions with maximum fidelity, the transfer function must allow the system to correctly interpret the measurements, infer the correct meaning, and enable the virtual actions to affect the simulation appropriately. 

This component is close to the \textit{control symmetry} dimension of the revised FIFA~\cite{mcmahan2016.UncannyInteration}. It is described to only depend on the transfer function symmetry where the system's transfer function is contrasted with a theoretical transfer function from reality. While we agree that no transfer function that would correspond to this system component exists in reality, we consider matching the virtual replication of the user actions with its counterpart of the reference interaction more practical than constructing a theoretical dummy.

\begin{wrapfigure}{l}{0.12\columnwidth}
    \includegraphics[width=0.169\columnwidth]{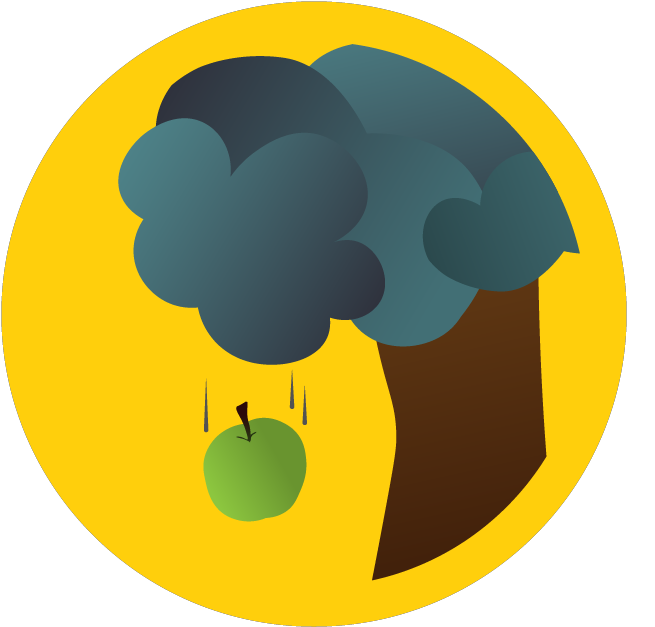}
\end{wrapfigure}    
\subsubsection{\textbf{Simulation Fidelity}} 
Simulation fidelity is the degree of exactness with which a virtual environment resembles the characteristics of the reference interaction's world and adequately reacts to the user's actions. 
In contrast to the next component, \textit{rendering fidelity}, it is irrelevant for simulation fidelity how the VE is rendered as output. 
While simulation fidelity considers the environment with its characteristics, rendering fidelity converts those characteristics through a dynamic process to the output devices (i.e., rendering pipeline).

This component concerns all elements of the simulated environment, including objects, agents, physics, and any other entity or logic from the reference interaction's world. To help categorize these aspects and to avoid simulation fidelity being a ``black box'', we define four subcomponents: presentational fidelity, behavioral fidelity, physical fidelity, and scenario fidelity. 

\textit{Presentational fidelity} is the degree of exactness with which the presentation of the simulated world resembles the reference world. 
Properties that can affect presentational fidelity include an object's mesh, colors, materials, scent, surface, reflectivity, and level of detail. For example, a low-poly presentation of an apple with a simple color-based material would have lower presentational fidelity to a real-world apple than a laser-scanned presentation with a photorealistic texture-based material. 

\textit{Behavioral fidelity} is the degree of exactness with which the behaviors of agents (e.g., virtual humans) within the simulation resemble the behaviors of their counterparts in the reference interaction's world. The fidelity of these behaviors depends on three aspects of the agent: its perceptual model, cognitive model, and motor model \cite{mcmahan2018.SystemFidelity}. The perceptual model defines what information is made available to the agent, which can vary from all information about the state of the world (i.e., fully observable) to a subset of information (i.e., partially observable) \cite{Russell_Norvig2020}. The cognitive model defines how the agent will generate actions based on the information provided by the perceptual model, such as simple reflexive responses or more complex decisions based on prior events and the agent's goals. The motor model defines what observable actions the agent can control, such as body movements, facial expressions, speech, or manipulating objects within the environment. Hence, a simple reflex agent that chooses actions from a small set based on current, partially observable information would have less behavioral fidelity than a goal-based agent that chooses actions from a large set of possibilities given fully observable information about the simulation. 

\textit{Physical fidelity} is the degree of exactness with which the physics of the simulation resembles the physics of the reference interaction's world. Physical fidelity can be affected by shapes, mass, distribution of mass, drag, gravity, or whether collisions are calculated discretely or continuously. For example, when skipping a stone over water, a high-fidelity system would simulate how the stone bounces due to the surface tension, travels a realistic trajectory, and creates ripples on the water's surface. In contrast, a stone with low physical fidelity (such as a box collider, a default mass of 1kg, and simple force estimations) would simply sink to the ground.

\textit{Scenario fidelity} is the degree of exactness with which the simulated situation resembles the situation of the reference interaction. While presentational and behavioral fidelity focus on individual objects or agents, scenario fidelity focuses on the holistic aspects of the simulation. This involves the spatial relationships among objects (e.g., a chair is usually placed under a table and not on top of it), their semantic relationships (e.g., an indoor room usually has a ceiling), and their logical relationships (e.g., flipping a light switch turns the light on or off). For example, a VE with a flat ground plane and no vegetation would provide less scenario fidelity to hiking through a mountain forest than an environment with uneven terrain and numerous trees.

As an example of different simulation fidelity, consider the reference interaction of playing tennis against a friend at your local park. A low-fidelity implementation may use low-poly representations of the tennis ball, rackets, and net with a basic virtual agent as the opponent that only reacts to the user hitting the ball, which follows a simple trajectory. On the other hand, a high-fidelity implementation would use high-poly representations in an outdoor park environment surrounding the tennis court, a virtual human driven by its own goals and capable of social interactions (e.g., congratulatory comments), and advanced physics that allows the user to apply topspin when hitting the ball. To virtually reproduce interactions with maximum fidelity, the simulation must replicate the reference world identically, including its objects, agents, physics, and situation.

The first part of the definition is based on \citet{mcmahan2011exploring}. This component closely links to the concept of simulation fidelity by \citet{nilsson_DecreasedFidelity_2017}. They accept the definition by \citet{mcmahan2011exploring} and further attribute simulation fidelity to ``the realism of the models forming the basis for the generation of the VE (e.g., geometric, lighting, or physical models).'' While we largely agree, we extend the scope of simulation fidelity by the system's response to the user's input.
This is also reflected in the concept of functional fidelity by \citet{alexander2005gamingTraining}, which requires in a training context that ``the simulation acts like the operational equipment in reacting to the tasks executed by the trainee.'' However, in this definition, the characteristics of the VE independent of the task execution are disregarded.  Further, we adopt and expand the subcategories \textit{attribute, behavioral, and physical coherence} by \citet{mcmahan2018.SystemFidelity} as well as \textit{physical and functional characteristics} by \citet{hays1988.SimulationFidelityTraining}.

\begin{wrapfigure}{l}{0.12\columnwidth}
    \includegraphics[width=0.169\columnwidth]{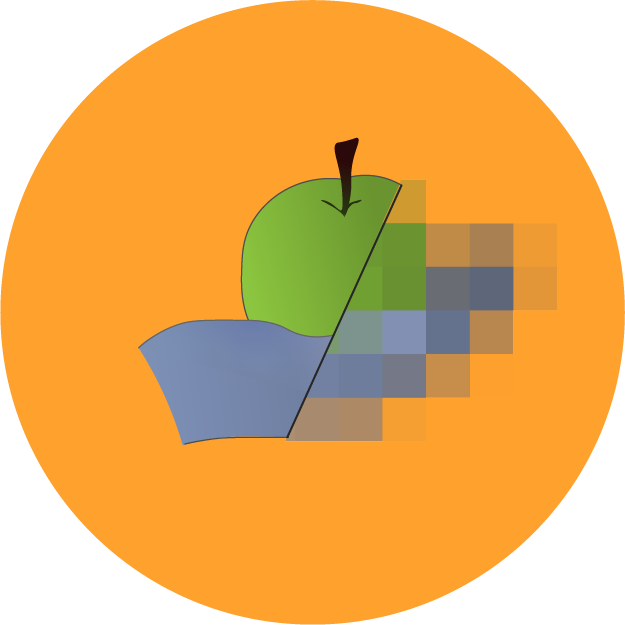}
\end{wrapfigure}
\subsubsection{\textbf{Rendering Fidelity}} 
Rendering fidelity is the degree of exactness with which the output content generated by the computer resembles what would be presented to the user in the reference interaction.
In contrast to the next component, \textit{display fidelity}, it is irrelevant for rendering fidelity whether the output devices can display the rendered output (accurately) and make it perceptible to the user.

This component involves any modality, including visual, auditory, haptic, olfactory, gustatory, and vestibular stimuli. The sensory stimuli must be rendered for any output device of the system. Systems with a visual display must render graphical images that can differ in their resolution, aspect ratio, framerate, visual style, antialiasing, texture resolution, detail of height maps, shadow, specular effects, and much more. Similarly, the sensory stimuli for any other kind of display must be rendered considering the respective parameters affecting the level of fidelity, such as audio, haptics, etc. (see \textit{display fidelity}). 
The rendering parameters do not necessarily correspond to the parameters of the simulation or the display; e.g., the rendered audio might compress the audio sources from the simulation and still have a higher resolution than what the earphones can display. 

As an example, for the sound of a virtual human walking, a low-fidelity solution would play a loop of the same footstep recording with strong compression. In contrast, a high-fidelity solution would synchronize the timing with the foot movement, adjust the volume, pitch, reverberation, and direction to the user's position, and dynamically blend high-resolution sound samples matching the floor material and the impact of the foot. 
To virtually reproduce interactions with maximum fidelity, the rendered output for all modalities must be indistinguishable from what the user would perceive in the reference interaction.

This component is related to the definition of display fidelity by \citet{mcmahan_DisplayIntFidelity_2012}: ``the objective degree of exactness with which real-world sensory stimuli are reproduced.'' In our model, we divide this aspect into rendering and display fidelity to reflect the independence of calculating output from displaying it.

\begin{wrapfigure}{l}{0.12\columnwidth}
    \includegraphics[width=0.17\columnwidth]{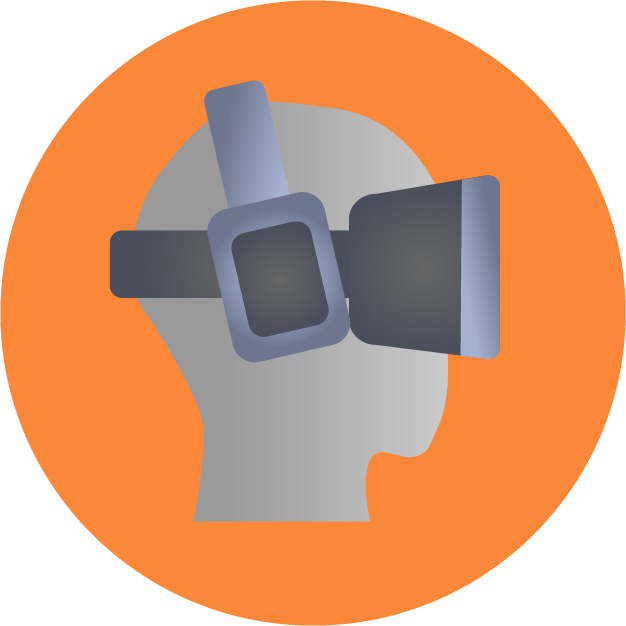}
\end{wrapfigure}
\subsubsection{\textbf{Display Fidelity}} 
\label{lab:DisplayFid}
Display fidelity is the degree of exactness with which the output devices reproduce the physical stimuli presented to the user in the reference interaction.
In contrast to the next component, \textit{perceptual fidelity}, it is irrelevant for display fidelity whether and how the user perceives the stimuli. 

This component covers displays that concern any modality, including visual, auditory, haptic, olfactory, gustatory, and vestibular displays. Since the alignment of output devices and human perception is highly complex, countless factors must be considered for display fidelity. For instance, concerning the graphical output of a head-mounted display, visual fidelity can be affected by the screen resolution, pixel density, field of view, refresh rate, contrast, and color depth, but also optical properties of the lenses and optical interferences such as god rays. The situation is again entirely different in a CAVE system with wall projections. With any modality, the properties of the displays might not match those of the renderings or what the user would be capable of perceiving, e.g., a screen might have a different resolution than the rendered image and the user's retina. 

As an example of haptic fidelity, considering the task of sawing through a plank, a low-fidelity solution would create static vibration feedback in a hand-held controller when moving the saw. In contrast, a high-fidelity solution would display force feedback that resembles the resistance of the wood and restricts the hand's lateral movement, as well as generate contact forces from the saw handle on the hand, render the gravitational pull of the heavy saw, and produce dynamic vibrations matching the jerky movement through the wood. 
To virtually reproduce interactions with maximum fidelity, the sensory stimuli displayed to the user address all senses and perfectly resemble the stimuli in the reference interaction. 

This component is related to various frameworks that detail display fidelity addressing a single sense, such as the Haptic Fidelity Framework by \citet{hapticfidelityframework}, components of visual display fidelity according to \citet{bowman2007.EnoughImmersion}, or dimensions of the Spatial Audio Questionnaire by \citet{lindau2014spatial}.

\begin{wrapfigure}{l}{0.12\columnwidth}
    \includegraphics[width=0.169\columnwidth]{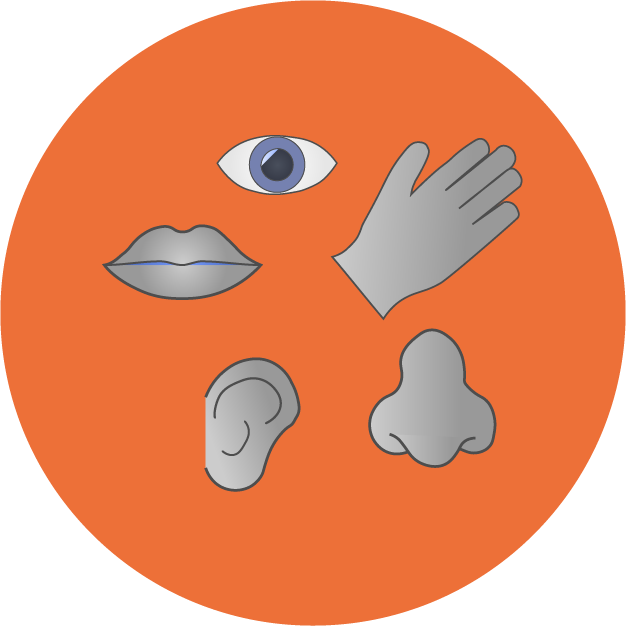}
\end{wrapfigure}
\subsubsection{\textbf{Perceptual Fidelity}} 
Perceptual fidelity is the degree of exactness with which the user's perception of the physical stimuli created by the system resembles how the user would perceive the reference interaction.
In contrast to the next component, \textit{experiential fidelity}, it is irrelevant for perceptual fidelity how users interpret the perceived stimuli, what meaning they assign to them, and what consequences they draw. 

This component concerns all sensory cues that the user registers with any sense: vision, audition, touch, smell, taste, proprioception, equilibrioception, nociception, etc.
The physical stimuli produced by the output devices are registered through sensory receptors. The information is transduced to and processed in the user's brain. The impressions from the different senses are integrated into one unified perception, called multimodal integration. We consider the interpretation of the stimuli already part of the user's experience and, therefore, included in experiential fidelity. 
We assume that maximum display fidelity inevitably leads to maximum perceptual fidelity because the human sensory system cannot identify the origin of a physical stimulus, whether the real world or a simulation generates it. With an imperfect display, however, perceptual fidelity may deviate. Mechanisms in human perception can compensate for display deficiencies, enabling phenomena such as illusions, biases, or sensory substitution. Thus, high perceptual fidelity may be achieved even without high display fidelity in some cases.
Further, the qualia of perception can vary immensely between individuals, which is why perceptual fidelity must be assessed on a user-by-user basis. Consider a color-blind user who experiences the real world in shades of grey. A black-and-white scene would look more realistic to that user than to users with full-color vision, thus yielding higher perceptual fidelity. 
Considering future possibilities of perception through direct neural manipulation without the respective receptors being stimulated, our understanding of this component remains the same: How closely does the user's perception correspond to the reference? However, the restriction to simulation through physical cues must then be omitted. 

In the example of someone touching different locations on the user's back, a solution with low display fidelity, such as a haptic vest with low-resolution actuators, would still result in high perceptual fidelity, as the two-point discrimination of skin receptors on the back is relatively poor \cite{lederman2009haptic}. A low-fidelity solution would have the same display resolution on the user's hand as the receptors are more sensitive to local variations here. Perceptual fidelity is also low in this example if the user only perceives visual cues and feels no tactile sensation. 
To virtually reproduce interactions with maximum fidelity, the stimuli by the output devices must evoke the same sensation for all senses and create the same perception in the user's brain as in the reference interaction.

\begin{wrapfigure}{l}{0.12\columnwidth}
    \includegraphics[width=0.169\columnwidth]{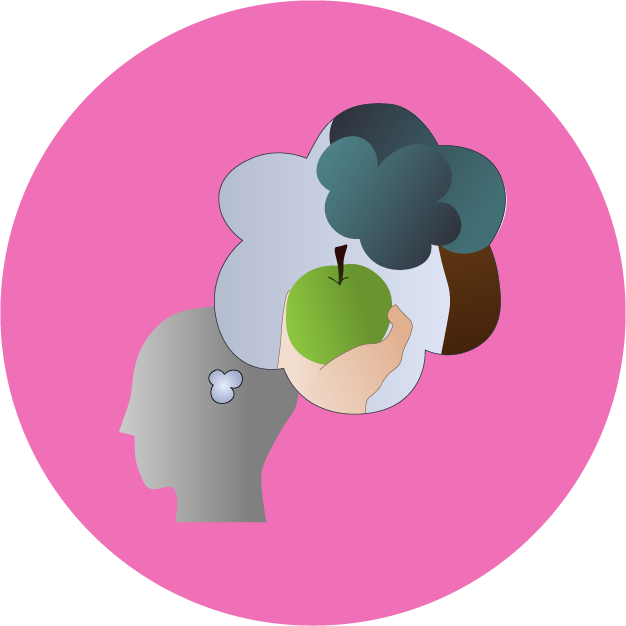}
\end{wrapfigure}
\subsubsection{\textbf{Experiential Fidelity}}
Experiential fidelity is the degree of exactness with which the user's experience of the simulated interaction resembles how the user would experience the reference interaction. When comparing to the real world, this is often referred to as \textit{perceived realism}.
In contrast to the next component, \textit{action fidelity}, it is irrelevant for experiential fidelity how the user reacts to the simulation.

This component is based on how faithful to the reference interaction the user considers their own actions, how the observed world behaves, and how it can be perceived. 
Experiential fidelity is often the ultimate objective when optimizing any other component, as the perceived authenticity can determine the subjective quality of a simulation. In other use cases, however, it can be subordinate, as in a technical proof of concept or a training situation that must prioritize action or simulation fidelity, whether experienced as faithful or not. Curiously, an objectively highly realistic system is not necessarily experienced as such, as discussed in Section~\ref{ssec:normativepower}.
Various factors affect experiential fidelity, such as individual differences in perception and judgment, since systems are often not optimized for one specific user and use case but are designed to be versatile and adaptable. Also, suspension of disbelief---or lack thereof---can substantially impact perceived fidelity, e.g., due to different assumptions and expectations, distractions from other realities, and the credibility or plausibility of the simulation. Also, unconscious effects need to be considered. For example, users can experience a higher cognitive load even for unnoticeable manipulations with redirected walking techniques~\cite{bruder2015.CognitiveDemandsRedirWalk}. Furthermore, multisensory integration can influence how users interpret conflicting stimuli, which can be demonstrated with phenomena such as the McGurk effect~\cite{tiippana2014.McGurkEffect}.
Another critical factor is the awareness of the experience being simulated or the memory of having entered a simulation.

\renewcommand{\arraystretch}{1.5}
\begin{table*} [ht!]
    \caption{Selected frameworks, models, and instruments that include details specialized on a certain aspect of fidelity, listed with its original designation. The references are grouped by their correspondence to the \model\. This does not represent an exhaustive list of all relevant, prior literature, nor does it comprehensively address all subcomponents.}
    \Description{TBD}
    \label{tab:signpost}
    \rowcolors{0}{white}{lightgray!20}
    \begin{tabular}{m{.2\textwidth}|m{.7\textwidth}}
        \textbf{\model\ components} & \textbf{Literature covering this component} \\
        \hline
        
        Action Fidelity & \textit{Biomechanical Symmetry}~\cite{mcmahan2011exploring} (only active motions), \textit{Motion Realism}~\cite{Rogers2022.suchwow} (only in games)\\
        
        Detection Fidelity & \textit{Input Veracity / System Appropriateness}~\cite{mcmahan2011exploring}, \textit{Tracking System Fidelity}~\cite{al2022framework} \\
        
        Transfer Fidelity & \textit{Control Symmetry}~\cite{mcmahan2011exploring} \\
        
        Simulation Fidelity & \textit{Simulation System Fidelity}~\cite{al2022framework}, \textit{Physics Realism / Avatar Realism}~\cite{Rogers2022.suchwow} (only in games), \textit{Functional Fidelity}~\cite{alexander2005gamingTraining}, Naive Physics~\cite{jacob2008reality} \\
                
        Rendering Fidelity & \textit{Software}~\cite{hapticfidelityframework} (only haptics), \textit{Visual/Graphic \& Auditory Realism}~\cite{Rogers2022.suchwow} (only in games) \\
        
        Display Fidelity & \textit{Hardware}~\cite{hapticfidelityframework} (only haptics), \textit{Spatial Audio Quality}~\cite{lindau2014spatial} (only audio), \textit{Visual \& Auditory \& Haptic System Fidelity}~\cite{al2022framework}, \textit{Device Realism}~\cite{Rogers2022.suchwow} (only in games)\\
        
        Perceptual Fidelity & \textit{Sensing}~\cite{hapticfidelityframework} (only haptics), \textit{Spatial Audio Quality}~\cite{lindau2014spatial} (only audio), \textit{Sensory Realism}~\cite{Rogers2022.suchwow} (only in games), Body Awareness and Skills~\cite{jacob2008reality} \\
        
        Experiential Fidelity & \textit{Experiential Fidelity}~\cite{experientialfidelity}, \textit{Player Response Realism}~\cite{Rogers2022.suchwow} (only in games), \textit{Presence}~\cite{slater1997framework, skarbez2017.Presence}, \textit{Psychological Fidelity}~\cite{alexander2005gamingTraining}, \textit{Haptic Experience (HX)}~\cite{kim2020defining}, Environment \& Social Awareness and Skills~\cite{jacob2008reality} \\

        \hline
        Input Fidelity & \textit{Action Fidelity}~\cite{stoffregen2003nature}\\
        Output Fidelity & \textit{Physical Fidelity}~\cite{alexander2005gamingTraining}, \textit{Digital Sensory System Fidelity}~\cite{al2022framework}\\
        System Fidelity & \textit{System Fidelity}~\cite{mcmahan2018.SystemFidelity}\\

    \end{tabular}
\end{table*}

For example, when interacting with highly but not perfectly realistic virtual humans, the experiential fidelity has often been found to be low due to the uncanny valley~\cite{Mori1970.UncannyValley}, despite high rendering and display fidelity. On the other hand, even with low display fidelity, it is possible to achieve high experiential fidelity, e.g., Valve's experience of drawing a longbow with only vibration and sound feedback using a hand-held controller.\footnote{The Lab (\url{https://steamcommunity.com/app/450390?}, last access: 2023-08-15). Valve Corporation, 2016.} Another example is the rubber hand illusion and its virtual replication~\cite{Ma2015.RHI}, which demonstrates how users can have strong body ownership and perceive haptic sensations with high perceived realism despite only sensing visual cues.
To virtually reproduce interactions with maximum fidelity, the interaction with the system must convince the user to experience the reference interaction, not a simulation. This would be the equivalent of a successful Turing Test for VR interactions, as proposed by \citet{stoffregen2003nature}.

This component is related to concepts such as presence, immersion, coherence, or body ownership and is assessed in several corresponding questionnaires~\cite{skarbez2017.Presence}. We discuss this in more detail in Section~\ref{sec:DisRelatedConstructs}.

\subsection{Dedicated Literature on Subcomponents}
With these eight distinct components, the \model\ illustrates what design decisions influence the fidelity and characteristics of VR systems enabling immersive interactions. When zooming out, the model provides an umbrella framework. When zooming in, further specialized models are needed to investigate the underlying complexity of the components. In multi-modal simulations, fidelity has countless detailed determinants that can be finely dissected as required. For example, when looking at the wrinkles of a virtual human, we can go further down in the component hierarchy: interaction fidelity → simulation fidelity → presentational fidelity → 3D model → skin texture → height map → resolution. But while skin characteristics can be rendered visually, they can also be rendered haptically: interaction fidelity → display fidelity → haptics.
The Haptic Fidelity Framework~\cite{hapticfidelityframework} is a good example that illustrates the complexity of one of the modalities of display fidelity. The framework comprises 14 distinct criteria defining just this one output modality. 

In the interest of this model's simplicity, we refrained from further detailing the included components. Instead, the work builds on various rich and informative works we refer to in \autoref{tab:signpost} as a signpost. It lists related frameworks and models from the literature that tie into the components of the \model. They specialize in one or a few aspects and provide in-depth information as needed. While the referred works provide further details concerning a component, they are not necessarily in exact correspondence with the component. 
Beyond the works listed, we encourage the HCI and VR community to devise further dedicated frameworks and measurement instruments to fill the current gaps. For instance, regarding simulation fidelity, the broad range of influences is not yet covered adequately by any specialized framework. Further, while the FIFA framework~\cite{mcmahan2016.UncannyInteration} allows a detailed analysis of body movements as a part of input fidelity, other elements of user actions and states are disregarded, such as speech, gaze, or body temperature.

\begin{figure*}[t]
  \centering
  \includegraphics[width=\linewidth]{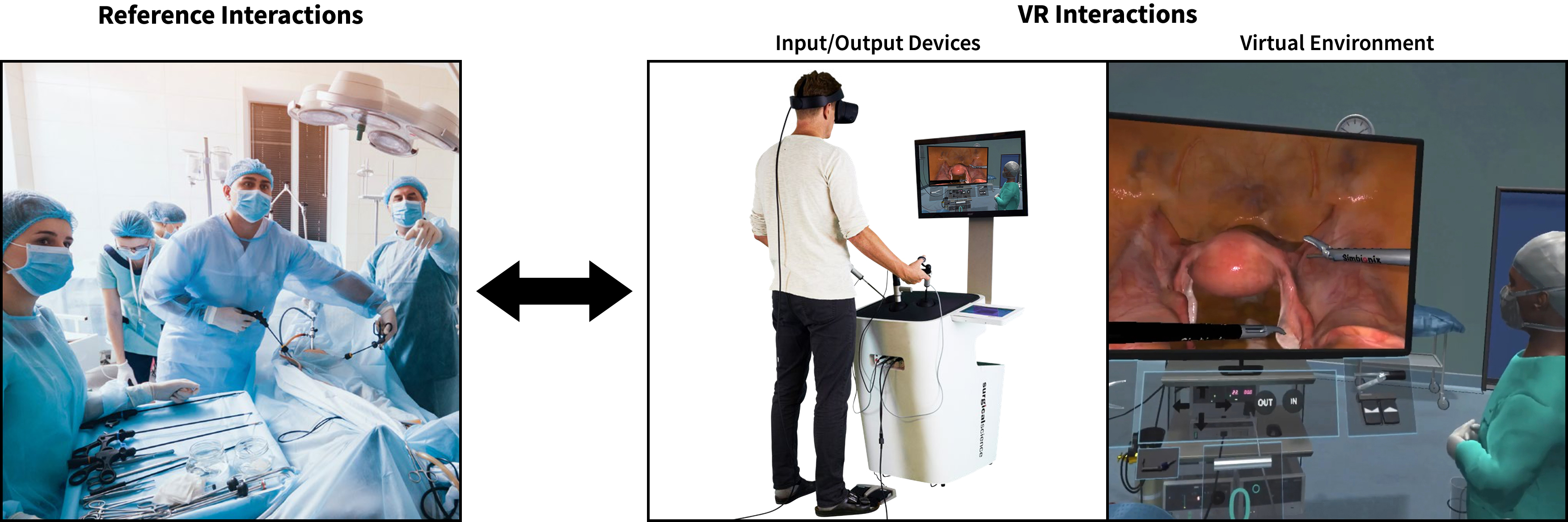}
  \caption{(left) The reference interactions in the real world to be simulated: A surgeon and his team performing a laparoscopy. (middle) The user operating the VR training system. (right) The user's view inside the VE: the operating room and a virtual screen with the laparoscopic camera view. \textit{Images by Surgical Science, modified by the authors.}}
  \Description{Reference interaction on the left: A photograph depicts a team of surgeons operating with surgical instruments and equipment visible. In the center, representing the input/output devices, a man is wearing a VR headset and manipulating controls on a simulation device. The screen in front of him displays the procedure, mirroring his actions. The rightmost image shows the virtual surgery environment. A virtual endoscopic view of a surgical procedure is depicted on a screen, with medical instruments interacting with virtual tissues. A virtual surgeon is standing next to it.}
  \label{fig:lapsim}
\end{figure*}

\section{Examples of Applying the Model}
\label{sec:examples}
Let us look at three diverse examples to bring the theory to life. In this section, we walk you through the analysis process of three use cases with different types of reference interactions. Example 1 demonstrates how we can use the \model\ to evaluate how realistic a training system for surgeons is. The goal is to come as close as possible to real surgery to practice under safe conditions. Therefore, high interaction fidelity is essential for acquiring skills in VR so they can be transferred to real-life surgery.
In contrast, Example 2 outlines skill training in which low-fidelity elements are helpful. Here, we assess the realism of a game for learning to juggle. The goal is to provide deliberately low interaction fidelity for effective training, accompanied by empirical research~\cite{adolf2019.Juggling}. 
Example 3 shows how the \model\ can be applied to fictional reference interactions. We discuss the fidelity of a VR game from the \textit{Star Wars} universe.

The authors conducted these exemplary assessments. They are subjective in nature, thus contestable. The examples illustrate how a complete evaluation based on the model could be performed. However, the outcome can differ depending on the analysis goals, context, and individual perspective. In our experience, disagreeing with an assessment and justifying the opposing opinion already provides a deeper understanding. Therefore, we encourage reasoned disagreement with our evaluation.

\subsection{Example 1: Surgical Training}

The LAPSIM\textsuperscript{\textregistered} by Surgical Science\footnote{\url{https://surgicalscience.com/simulators/lapsim/}} is a commercial surgical training simulator for laparoscopic interventions. The system offers medical simulation training for effective and patient‐safe training of surgical competence that can be transferred to the real operating room.\footnote{Overview of the system \url{https://www.youtube.com/watch?v=7wlIBBm1RXU}} Laparoscopic surgery is a technique in which short, narrow tubes are inserted into the abdomen through small incisions. Long, narrow instruments are inserted and used to manipulate, cut, and sew tissue, as shown in \autoref{fig:lapsim} (left). The LAPSIM is an advanced simulator with detailed graphics and haptic feedback to train these procedures. The system consists of a custom input device with two laparoscopic grips with precise tracking through a wire system and a third grip to control the camera position inside the virtual patient's body. The grips offer accurate haptic feedback for soft tissue and hard surfaces, such as bones, with force feedback delivered through the wire system. For the experience of being in a virtual operating room, the system is equipped with an Oculus Rift headset with outside-in head tracking. The Oculus controllers are not used. The virtual scene consists of a patient, a monitor with a view of the laparoscopic camera, and assisting surgery staff, e.g., nurses.

We now go through the loop clockwise, starting with the user's actions. 
The user interacts with the LAPSIM through custom laparoscopic grips. As a result, the user performs actions with their body that match well with real laparoscopic surgery, particularly the movements of the arms, fingers, and one hand. However, the other hand and the body posture differ as users stand comfortably in front of the LAPSIM while in the real surgery, they must lean in, as can be compared in \autoref{fig:lapsim}. Still, \textit{action fidelity} can be considered high. The actions' detection is realized with a wire system precisely measuring the motions in four DoF (three rotational and insertion depth) of the laparoscopic grips---the same as in the reference. The system provides very high \textit{detection fidelity} of the surgical tools. However, finger, arm, other body motions and the voice are not captured, decreasing \textit{detection fidelity}. The system transfers the measurements into appropriate positions and orientations of the virtual grips. In addition, the system uses inverse kinematics to estimate hand and arm motions based on the end position of the grips. This gives the system high \textit{transfer fidelity}. 

\begin{figure*}[ht]
    \centering
    \includegraphics[width=\linewidth]{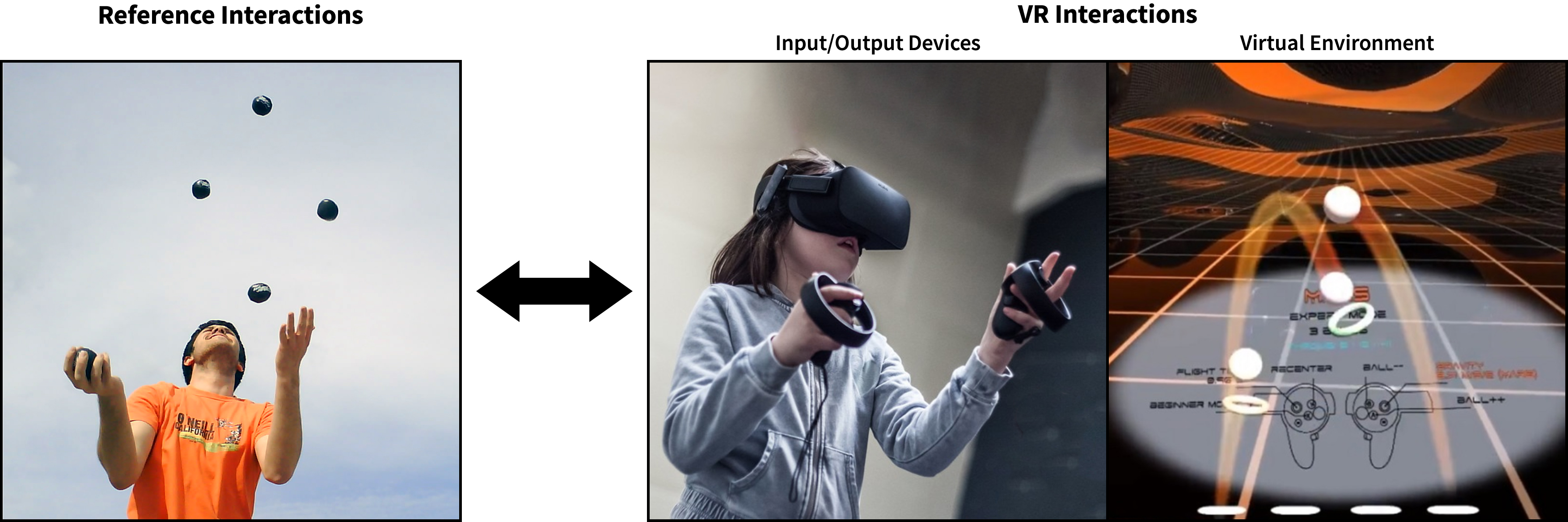}
    \caption{(left) The reference interaction of a person juggling with balls. (middle) A user juggling in the VR simulation using a VR headset and controllers. (right) The user's view within the VE: three balls are being thrown in a cascade pattern in a Mars-inspired environment. The two rings represent the user's hands. \textit{Left image by Loris Bottello (CC~BY-NC-SA~2.0, modified). Middle and right images (modified) from \cite{adolf2019.Juggling}}.}
    \Description{On the left, a person is juggling four balls outdoors, looking upwards as they catch and throw them in the air. In the middle, a child uses a VR headset and hand controllers indoors to juggle virtually. On the right, the virtual environment shows a juggling simulation, with a first-person perspective of hands and balls in motion, alongside interface elements that provide feedback or instructions.}
    \label{fig:juggling}
\end{figure*}

Based on the input, the LAPSIM simulates tissue properties very accurately, such as its softness or response when cutting it. When blood vessels get damaged, the system simulates bleeding abstractly in the form of blood spilling out but not flowing anywhere. Outside the surgical site, the behavior of the surgical staff is simulated quite well as they perform relevant tasks in the operating room and react to the progress of the surgery. However, their animations seem sluggish and unrealistic. Therefore, \textit{simulation fidelity} is in some aspects high, in others low. 
The VE in the LAPSIM can be considered as two separate parts: (1) the operating room the user is standing in and (2) the surgical site inside the virtual patient's body, which is displayed on a virtual screen within the operating room. The in-body view is rendered with highly detailed textures, reflections, and lighting for the internal tissue. On the other hand, the operating room and staff are rendered in a cartoon style that does not reflect the real environment well. Therefore, \textit{rendering fidelity} is in parts high and medium. 

The laparoscopic grips display the haptic feedback precisely by considering tissue softness and resistance. This aspect of \textit{display fidelity} is high. In contrast, the visual and auditory output is displayed on the Oculus Rift headset, which differs from what is perceivable in the real world due to constraints such as a limited field of view or screen resolution. Olfactory cues are missing. Concerning modalities other than haptics, \textit{display fidelity} is medium-low. 
Limited by each modality's display fidelity, the user's perception is equally restricted compared to reality. Hence, \textit{perceptual fidelity} is in parts medium-low or high. 

As a user's experience is highly subjective, we conducted an informal interview with an experienced surgeon to assess \textit{experiential fidelity}. He has practiced with the LAPSIM and frequently conducts this type of surgery. 
The surgeon described the haptic feedback as extremely close to the real world, with a high contribution to the experience, as this is the focus of the intervention. On the other hand, the animations were described as not convincing. Some surgery procedures are missing in the LAPSIM, such as changing the physical instruments by completely pulling them out and inserting a new instrument. 
Overall, the surgeon assessed the system to have high realism as it comes close to the experience of a real laparoscopic surgery with a match of 70\% to 80\%.

In conclusion, the LAPSIM provides high fidelity for aspects that are most important for the training of hand-eye coordination, surgical procedures, and the development of manual dexterity. The system provides medium to low fidelity for less relevant aspects, such as staff animations and environmental graphics. However, this suffices for the training purpose, as scientific validation confirmed. Studies have shown that skills trained in LAPSIM can be successfully transferred to real surgery~\cite{calatayud2010warm, cosman2007skills}.
Our detailed assessment of LAPSIM's interaction fidelity identifies the strengths and weaknesses of the system, confirms adequate prioritization in its development, and shows opportunities to improve realism further. The example demonstrates that thoughtful interaction and system design can help achieve the purpose of a VR system without improving fidelity in every aspect.

\subsection{Example 2: Learning to Juggle}
Juggling with three or more balls is a complex activity that can be challenging to learn. There is a steep learning curve as you either throw and catch the balls with the correct timing, or they will fall to the ground repeatedly. To make learning to juggle easier, a virtual simulation can deliberately deviate from the reference interaction in the real world, thus lowering interaction fidelity. In this example, we analyze the VR software \textit{Planet Juggle} by Benjamin Outram,\footnote{Description, video trailer, and free download of \textit{Planet Juggle} for the Oculus Quest and Rift at \url{https://www.benjaminoutram.com/planet-juggle}, last accessed 2023-08-15} which provides various features to facilitate a gentler learning process of juggling movements, such as the cascade pattern. 
For instance, the user can activate the following assistive features. Slow motion allows the user to practice and internalize movement sequences without getting hectic. When touching, the balls can snap to the hand to make catching easier. Visual indicators can show the trail of the balls and preview where they will go to achieve the ideal cascade trajectory. The balls can always reach the ideal height independent of the throwing impulse. And there is background music that helps get the ideal rhythm. When getting more confident, the user can turn off the features for closer-to-reality training. 

Again, we now systematically look at the interactions' fidelity along the loop when all assistive features are activated. 
The 1-to-1 mapping of the motions within the 3D space would allow high correspondence of the body movements to the movements of real juggling. However, the grasping and releasing actions differ since the user controls the virtual balls with the trigger buttons of the hand-held Oculus controllers instead of physical balls. For holding the controllers comfortably, the hands are rotated inwards for virtual juggling, other than in real juggling, where the palms must face up to catch a ball. Further, supportive features such as snapping, slow motion, or automatic height make the system tolerant of discrepancies in hand position or rotation, throwing power, and bad timing. While this allows successful first steps in virtual juggling, the same movements would lead to failure in the real world, resulting in low \textit{action fidelity}. 
The detection of hand movements works precisely using the Oculus tracking systems. Other body movements are not registered, although less relevant to the activity. Therefore, we can assess high \textit{detection fidelity}. 
The system tolerates discrepancies in hand movements by making up for them with transfer functions. Although the user actions may differ from realistic juggling movements, the transformed virtual juggling actions closely correspond to the reference interaction, which leads to high \textit{transfer fidelity}.

\begin{figure*}[ht]
    \centering
    \includegraphics[width=\linewidth]{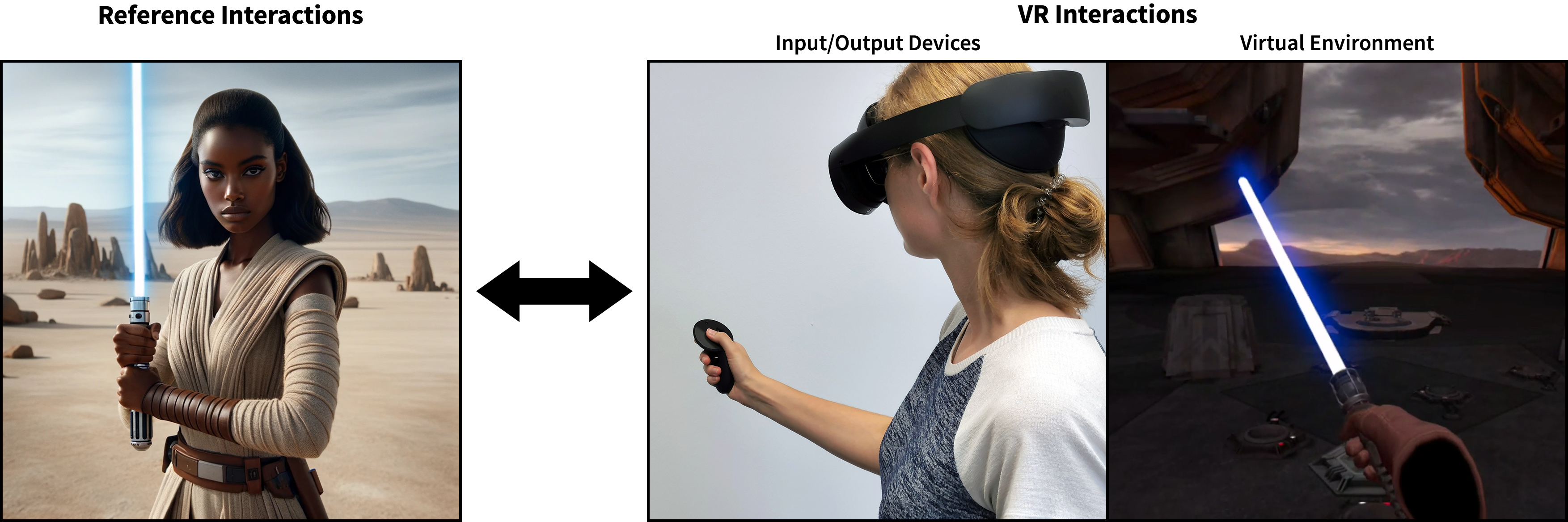}
    \caption{(left) A person holding a lightsaber as the reference to be simulated. (middle) A user with a headset and a controller using the virtual lightsaber. (right) The player's view in the VE. \textit{Left image generated with OpenAI DALL·E 3, right image captured from ``Vader Immortal: A Star Wars Series -- Episode III''.}}
    \Description{On the left is a photorealistic render of a woman holding a blue lightsaber, standing in a desert-like environment. In the middle is a photo of a woman wearing a Meta Quest Pro. She is extending her arm forward, holding a VR controller. On the right is a digital render with the first-person perspective of the virtual environment, with the user's hand holding a lightsaber identical to the one in the left image, in the same pose as the person in the middle image. A dark desert landscape is in the background.}
    \label{fig:lightsaber}
\end{figure*}

The lower gravity leading to the slow motion effect fits the narrative of \textit{Planet Juggle} as the user is supposed to juggle on the Moon or Neptune. Disregarding side effects such as the dreadful death from cold, the simulation's physics is generally sophisticated regarding ball behavior. However, our reference interaction for this comparison is real juggling on Earth. Features such as slow motion or visual trajectory indicators limit realism. Particularly, the strong predetermination of where the balls can travel and land, such as the spatial confinement to a plane disregarding the third dimension away from the user, leads to low \textit{simulation fidelity}. 
Moving on to output fidelity, the system renders an abstract environment with grids on the ground, geometrical landscapes, and glossy, highly reflective surfaces. Rhythmic music is generated to influence the juggling tempo, and the balls create abstract sounds when touching the hands.
All this can be considered low \textit{rendering fidelity}. 
The visual quality depends on the Oculus hardware used, but even with the superior Rift headset, the visual impression can be clearly distinguished from what the user could see in reality. Similarly, haptic fidelity is limited because the system only provides continuous contact forces from holding the controller and abstract vibrations when touching a ball. Overall, \textit{display fidelity} is medium to low. 
The controller's vibration feedback acts as a sensory substitution. The user cannot feel the momentum and weight of the falling ball, but the vibration intensity is calculated from these parameters. The haptic cues can inform the user about the ball's properties and the impact of catching. Similarly, audible cues represent the ball's force on the user's hand. Consequently, \textit{perceptual fidelity} can be considered higher than display fidelity, with medium fidelity.

Finally, to assess \textit{experiential fidelity}, we tested the simulator ourselves with a think-aloud approach. The app gave a juggling novice the impression of quickly acquiring juggling skills as the movements for the cascade pattern were quickly performed. He felt competent to virtually juggle after a few minutes, thanks to the assistive features. On the other hand, transferring these skills to the real world was a completely different story, as too many aspects differ from juggling with real balls. 
Another of the authors, proficient in juggling, struggled with the skill transfer in the other direction. At first, he could not accomplish a stable juggling pattern because it felt so different from what he was used to. Mainly the haptics were found to be too different as the soft vibration gave no impression of an impact. 
Also, the lack of catching and releasing the balls with the hand felt unfamiliar. The author further missed the third dimension because the ball trajectory is restricted to a 2D plane. He disliked the assistive features in VR: ``I find reality much more `assistive.' I missed that in VR.''
Overall, the authors enjoyed the playful VR activity but did not get the feeling of real juggling. We assess \textit{experiential fidelity} subjectively as low. 

In summary, the simulator provides virtual aids to learn the principles of juggling by deliberately deviating from a highly realistic replication. The supportive features allow quick progress for virtual juggling and can help build muscle memory through authentic arm movements. A scientific evaluation of the app has explored how it can be integrated into learning to juggle with real balls~\cite{adolf2019.Juggling}. The study showed how users had more fun training with the VR simulation but struggled with transferring the acquired skills. Therefore, developers should carefully consider which fidelity aspects should maintain high realism to ensure skill transferability.

\subsection{Example 3: Lightsabers from \textit{Star Wars}}
In this third example, we turn to a reference interaction from a fictional narrative: using a lightsaber, the energy sword from the \textit{Star Wars} franchise. Specifically, we examine the interaction fidelity of using lightsabers in the VR game \textit{Vader Immortal: A Star Wars Series -- Episode III} by ILMxLAB.\footnote{Trailer and download at \url{https://www.oculus.com/experiences/quest/2426206484098337/}} The player can use the weapon in lightsaber duels, to block blaster bolts, hit enemies, and throw it at targets. It is operated with hand-held Oculus Touch controllers. 
But what do we compare the VR interactions to if nobody has ever held a real lightsaber? The depiction in Star Wars media that also informed the game's development seems a reasonable match. 
Even in the best case, a fidelity assessment is debatable as we currently lack objective measures. Our evaluation is even more contestable when we compare to a fictional reference interaction since that already leaves room for speculation and disagreement. Interpretations can vary depending on the canonical choice of Star Wars media, such as comics, TV shows, books, video games, or merchandise artifacts. For this reason, we focus on the original movies as a reference.
We invite all readers to question our stance and justify their proposal. 

Instead of a circular procedure, we look at the most striking characteristics of the interaction. 
The weapon functions in the virtual world much as you would expect it to. The player holds the lightsaber's hilt, wields it, and throws it as the plasma blade blocks attacks and cuts through any objects. However, most objects are not cut into pieces when the blade goes through them, and it hardly leaves a trace. Due to the simplistic game mechanics of the choreographed lightsaber duels, also the opponents do not always respond to being touched by the lightsaber; similarly, the player does not die instantly from being hit by an enemy's lightsaber. In this game, using a lightsaber has medium \textit{simulation fidelity}. 
The controllers as input devices are beneficial for \textit{action fidelity} because the player has a similar hand pose as when gripping the lightweight, slim saber hilt. Even the activation by pressing a button corresponds well. How the player holds the weapon and performs combat movements with six DoF contributes to overall high \textit{action fidelity}. It is sometimes lower, though. If the player holds the hilt with both hands, having two separate controllers limits action fidelity. It further suddenly decreases when there should be resistance from hitting another lightsaber, but the player's arm continues to move unhindered. 

This equally affects \textit{display fidelity} due to the lack of force feedback. Because the controllers only provide passive haptic feedback and vibrations, we assess this aspect as low-fidelity, even more so when virtually throwing the lightsaber but still gripping the controller. The headset's constraints on the visual impression reinforce this impression.
At the same time, \textit{transfer fidelity} can be considered high as the wrist's flicking motion and the grip button's release are interpreted as a throw, sending the weapon on a boomerang trajectory in an adequate direction. Generally, movements are recognized accurately, resulting in high \textit{detection fidelity}.
When activating the weapon, the player hears the iconic hissing sound. The pitch of the electric swoosh sound is higher when wielding the sword. Together with the convincing glowing effect of the blade, we thus assess \textit{rendering fidelity} as high. 
Given the limited skills of the testing author, we are comforted by the low nociceptual fidelity as there is no pain from getting hit. Instead, the view is overlaid with a red vignette, and we hear the avatar moan, which could be considered a decreasing factor for rendering fidelity. 
Apart from that, \textit{perceptual fidelity} is restricted by low display fidelity. 

Lastly, to inform our assessment of \textit{experiential fidelity}, we informally tested the game with a Star Wars enthusiast and someone who has only seen a few movies. The enthusiast was amazed at how much the lightsaber made him feel like a Jedi. It felt convincing, wielding a glowing, humming lightsaber in 3D space with high levels of embodiment, especially compared to playing the game series \textit{Star Wars: Jedi Knight} on the computer with a mouse and keyboard or using sticks as a child. Only the crude combat game mechanics detracted from the experience. 
The other tester, who did not care about Star Wars, also enjoyed handling the lightsaber. When asked if this is how he would expect a lightsaber to feel and behave, he was unsure as he never contemplated it. He was surprised that there was no air resistance or weight like from a metal longsword but then assumed: ``Probably, this is just what a lightsaber feels like.'' Furthermore, he was confused by the inconsistent controls: a blaster requires continued pressing of the grip buttons while the lightsaber only needs one short press. Also, the vibrations and the headset's visual limitations were described as irritating. Overall, both testers quickly forgot about the system's shortcomings and were immersed in the experience. We suggest a medium-high fidelity rating.

This example demonstrates that the reference interaction does not need to be based on the real world. As long as the original to be simulated is clearly defined, the \model\ can be applied meaningfully. Regarding experiential fidelity, users with no knowledge of the reference cannot make a competent comparison, but they still have their assumptions. Similarly, non-swimmers cannot compare virtual swimming to real experiences but can make an informed guess, assess the simulation's coherence and credibility, and form their impression intuitively. 
Moreover, the lightsaber interaction exemplifies how other factors (such as fun in gaming) should be prioritized over maximum fidelity. The game would hardly be playable if the weapon sliced every object and the self-avatar, when touching it with the deadly blade.

\section{Validation}
\label{sec:validation}
To polish and validate the \model, we conducted expert interviews with 14 established VR specialists from academia and the industry. The goal of this evaluation was to collect feedback and criticism, discuss the proposed terminology, and put the model to the test in terms of its applicability.  

\subsection{Method}
We conducted semi-structured, interactive interviews via Zoom or in person. We prepared an interview guideline to structure the conversations and make them comparable. The only variation between the interviews concerned the experts' example projects to which they applied the model during the interactive sessions. At least two of the authors were present for each session. We conducted the interviews between March and May 2023. The conversations took between 45 and 79 minutes, averaging 63 minutes. We recorded the sessions with the participants' consent and collected 14.75 hours of material. We performed a thematic analysis of the data to process and structure the findings. Two of the interviews could not be recorded due to a technical error, so we resorted to notes on these sessions.
After finalizing the manuscript, we asked all interviewees whether the interviews had been adequately summarized in the reported findings. One expert did not find the time, but everybody else confirmed that the text accurately reflects the conversations without any requests for changes or additions. 

\subsubsection{Sample}
We selected researchers, interaction designers, developers, and managers for their contributions to the VR field and their professional experience. We invited 19 selected candidates, of whom two could not find the time, and three did not respond. The diverse sample covered a broad range of VR-related research fields, backgrounds, and perspectives. Because we assured all participants that they would remain unidentifiable, we can only report broadly on the sample's composition. 

The sample included five full professors, three assistant or associate professors, and a lecturer from academia, as well as two senior professionals, a research scientist manager, a product owner, and a consultant from the industry. 
Among the experts were distinguished scientists who were awarded three IEEE VGTC VR Technical Achievement Awards, the IEEE VGTC VR Significant New Researcher Award, an IEEE VGTC VR Best Dissertation Honorable Mention, and various career awards. The sample also included three IEEE VGTC VR Academy members and an ACM Distinguished Scientist. 
The ten researchers in our sample had an average citation count of 7,043 and an average h-index of 35 as of 14 February 2024 according to Google Scholar. The h-index ranged from 11 to 69 and reflects the spectrum of our selection that includes young scientists publishing with a focus close to the \model\ and experts with decades of experience in VR research.
All participants primarily work in VR or HCI. The experts’ focus includes haptics, locomotion, interaction techniques, embodiment, computer graphics, visualization, presence, training and education, avatars, perception, medicine, collaboration, artificial intelligence, and multimodal interfaces. The variety of topics demonstrates how broadly the model can be applied in practice.
After analyzing the last interview, we invited three interviewees to join the author team and further contribute to this work, which they accepted.

\subsubsection{Interview Structure}
The interviews followed a semi-structured guideline based on a slide deck.
\begin{enumerate}
    \item \textbf{Working definition.} We defined the terms fidelity and realism in the context of our model.
    \item  \textbf{Task 1.} Without any previous biasing, we asked the experts to describe the fidelity of the interactions in one of their works and with their own words. Find more details below on how we selected the work. 
    \item \textbf{Introduction to the \model.} We explained the purpose of the model, the objective approach, the fidelity spectrum, the HCI loop as the foundation, and the components of the model integrated into the loop. We then elaborated on the single aspects with definitions and characteristics. 
    \item \textbf{Initial feedback.} We asked for first thoughts, any confusion or questions, criticism, and concerns regarding the terminology. In some conversations, this feedback was already raised during the presentation of the model. 
    \item \textbf{Task 2.} We suggested different ways to apply the model. Once more, we asked the expert to elaborate on the interaction fidelity of the same work as in task 1, only this time using the \model\ as a basis. We discussed it in more detail and asked follow-up questions as needed, such as identifying key aspects with prioritized fidelity, low-hanging fruit for improving fidelity, connections between aspects, etc. 
    \item \textbf{Conclusion.} Finally, we asked for the expert's overall evaluation of the model. Depending on the time left, we went into more detail regarding possible benefits, improvements, or research opportunities that the experts wanted to add. 
\end{enumerate}

For the two tasks of assessing interaction fidelity in a specific example, we selected one project, system, or user study of the expert. The interview partners were familiar with the work, so we could directly dive into the discussion. This approach also allowed us to authentically test the practical applicability of the \model\ in genuine use cases. By using the example in both tasks, we could compare how the experts approached the analysis either with or without the model providing structure. While we observed the differing levels of detail and evaluation strategies between the first and second task execution, the experts could experience how the model can facilitate the assessment of interaction fidelity. 

The works covered a wide variety of topics, including avatars, locomotion, haptics, perceptual illusions, presence, object manipulation, social interactions, embodiment, emotions, virtual environments, and training. To preserve the anonymity of our interview partners, we do not specify the chosen publications or projects. The works were required to link to the model, i.e., depend on at least two aspects of fidelity, aim for any kind of fidelity, or be inspired by an explicit reference interaction that has been reproduced.

\subsection{Findings}
In this section, we present the insights from the expert interviews. We refer to the experts from our sample with E01 to E14. 
Overall, the interviewed experts found the \model\ ``useful'' (E04), ``comprehensible'' (E07), ``sound'' (E02), and ``a meaningful contribution'' (E03). However, there was also reasoned criticism, opposing views, and a need for clarifying discussions. For example, some terms were described as ``not perfect'' (E11). Expert E10 was not convinced of our validation process' rigor, while E04 particularly liked it and found it ``very systematic''. E08 struggled with the HCI loop as a not entirely suitable basis but also found that ``some more modern reflection on the old interaction loop is definitely interesting.'' Each of our interview partners appreciated the \model\ as interesting and helpful.

The findings are structured by the identified themes: \textit{concept criticism}, \textit{terminology criticism}, \textit{applicability criticism}, \textit{application strategies}, and \textit{contributions}. The themes \textit{application traps} and \textit{patterns} were moved to the separate Sections~\ref{sec:traps} and \ref{sec:patterns}. Some critiques are addressed in the discussion section and, therefore, only mentioned here as a topic but not directly elaborated upon to reduce redundancy. Comments, misunderstandings, and concerns that were resolved in the interviews are not outlined. Similarly, any additional literature suggested by the experts was integrated into the Related Work Section but is not mentioned here separately. We directly adjusted the model and this paper according to the experts' suggestions without reporting before-and-after comparisons to avoid confusing readers with changes from an unknown earlier version. These unreported improvements primarily concern details of descriptions, definitions, framing, or illustrations.

\subsubsection{Theme: Concept Criticism}
The fundamental conceptualization of the \model, linking the fidelity aspects to the stages of the HCI loop, convinced most of our interview participants. E03 found that ``on a conceptual level, this is a very nice piece of work. I think it would be a meaningful contribution to how we think about and talk about virtual reality systems. [...] It's obvious a lot of thought went into this.'' One participant called the model ``a very high-dimensional matrix'' (E08).

However, there was also criticism on a conceptual level. The two most prominent critiques addressed the claim of objectifiability and the lack of quantification. Concerning the former, some participants doubted that all aspects of the model can be described objectively and struggled with the impracticality of ever determining an underlying ground truth. E14 hypothesized that system-related aspects (i.e., the bottom five) can only be assessed objectively, while the user-related aspects (i.e., the top three) only subjectively. We argue in the discussion (Section \ref{sec:discussion}) that all components can be assessed objectively and subjectively depending on the approach.
Concerning the latter, most experts indulged at some point in the interview in the idea of how helpful it would be to quantify interaction fidelity and its components: ``This scale from low to maximum fidelity: this is a scale from 0 to 100, and I want to have that value. It would be nice if I could assess that, of course.'' (E14). Enjoying the prospect of universal fidelity metrics and measurements ourselves, we explain in the discussion why this is not in the scope of this paper and might need many more years of systematic research. 

For some interviewees, the concept of fidelity was less intuitive than realism. They initially proposed reframing the model for comparisons to real-life interactions as the most common replication reference of VR simulations. After looking at practical examples with reference frames other than reality, most experts supported using the fidelity concept and preferred the more universal applicability. One of the FIFA authors criticized this limitation in their own work in hindsight: ``That was a weakness we had with [FIFA] that we were gauging everything with regards to the real world.'' 

One expert from our sample, E08, raised doubts that it makes sense to consider perceptual fidelity ``because we can't know---it's unknowable!'' 
Although it is currently challenging or impossible in many cases to determine what a user objectively perceives, there is a ground truth to it that has a higher or lower match to what the user would perceive in the reference interaction. With this, independent of its ascertainability, perceptual fidelity is a valid construct. Some interviewees pointed out that perception might even be measurable in the future if we develop more sophisticated neural interfaces. 
E08 continued, that ``obviously, nobody knows if perceptions are the same. That's the whole philosophical debate about qualia.'' For this reason, we recommend considering perceptual and experiential fidelity on an individual level. Only a user-by-user assessment can do justice to different personal abilities and characteristics. 
From a practical stance, E05 argued that while perceptual fidelity ``is user-dependent, it is not unknowable or unpredictable if you have enough information about the user.'' 

Another critique by E08 concerned the HCI loop as it is segmented into stages, while interactions in VR often occur with simultaneous input and output in parallel. Appreciating the debate about the loop's limited applicability for direct manipulation techniques due to the concurrence of input and output, we argue that it is nonetheless insightful to use the loop as a theoretical construct for abstracting and distinguishing the involved elements of the reciprocal VR interactions. Breaking down the permanent exchange of user and system can help understand the ongoing parallel processes while identifying isolated aspects relevant for analysis. 

Fundamentally, E08 was skeptical of operationalizing fidelity at all ``as it's obviously impossible to reproduce things.'' In our conversation, we agreed that perfectly replicating an original is extremely difficult and potentially impossible, especially regarding something as complex as bodily interaction with the world. However, using the \model\ to determine interaction fidelity is most helpful on the vast spectrum before reaching ``maximum'', i.e., to describe \textit{how} imperfectly something is reproduced. Realistically, we are nowhere close to the upper extreme of the spectrum with current technology. Until we achieve perfection, the model can help assess the degree of fidelity.

Several experts wondered how some aspects can be evaluated independently of others as they seem inseparably linked in practice. For example, rendering fidelity seems to be coupled to simulation fidelity on a computational level and to display fidelity on a hardware level. We agree that from the optimization viewpoint, the components should strongly depend on each other. Consequently, they are usually configured in combination by game engines and developers. Still, it can be beneficial to consider the aspects individually from an interaction design perspective. Consider a use case where a user cannot see the hair on his arm. It makes a difference if it is incorrectly represented in his avatar model, if it is not rendered due to missing height maps, or if the display resolution is too low for the hair to be recognizable.
Further, distinguishing the aspects can even be required from a technical point of view. For example, in cloud computing or when watching 360° videos, the rendered output can be displayed on HMDs with different screen resolutions, affecting rendering and display fidelity separately. 

As an addition to the model, E03 proposed integrating a ninth component along the middle axis of the loop: an element that moderates between the user's mental model (as part of experiential fidelity) and the system's data model (as part of simulation fidelity). We have not implemented the proposed extension of the \model\ for two reasons. First, there is no correspondence of the abstract concept of a mental model to any reference interaction, which is why the term ``fidelity'' does not apply. Second, the HCI loop has no equivalent component, and we prefer sticking to the original conceptual foundation.

\subsubsection{Theme: Terminology Criticism}
We asked the experts about the clarity and suitability of the chosen terms. Most interviewees had no concerns regarding any of the labels and appreciated how consistent they are. E02 said: ``I think the terms are perfectly clear as they are. And it's good that they're short, both in writing and when talking about it.'' Although many experts pointed out how introducing a defined vocabulary can be helpful for communication, others cautioned against conflicts with existing terms or definitions. In the literature and oral vocabulary, some terms are used differently from the \model\ understanding, and sometimes different terms are used for the same thing. For example, \textit{interaction fidelity} is coined more narrowly in the Framework of Interaction Fidelity Analysis \cite{mcmahan2016.UncannyInteration}. One of its co-authors, who is part of our sample, commented on the redefinition: ``In our previous work, we used [interaction fidelity] almost exclusively for the input side. But it was always a problem. [...] That will be a battle you'll have to fight to get the term to mean more. But it makes sense overall.''

Furthermore, there was criticism on a few specific terms, mainly by E11. We thoroughly discussed the critique with the experts and among the authors. In the end, we chose the term that seemed most suitable. For example, we changed \textit{sensory} to \textit{perception fidelity} to avoid confusing it with system sensors, and later from \textit{perception} to \textit{perceptual fidelity} for linguistic reasons and consistency with \textit{experiential fidelity}. 
We also changed the previous term \textit{feedback fidelity} to \textit{output fidelity} for higher symmetry with input fidelity and a better match with the component \textit{output devices} of the HCI loop. Moreover, the term \textit{feedback} implies a reaction of the system to user input. Hence, in a non-interactive $360^{\circ}$ movie without user input, there can be no feedback, but the system can still give output.

Other alternative labels were suggested in the interviews but not adopted after careful consideration. 
For example, \textit{tracking fidelity} was suggested instead of \textit{detection fidelity}, though this term implies including the transfer component. Alternatively, it was proposed to be called \textit{human sensory fidelity} while \textit{technical sensory fidelity} would replace \textit{perceptual fidelity}. However, this might facilitate misunderstandings and would be less concise.
Further, the label \textit{action fidelity} was criticized for not encompassing passive states and uncontrolled behavior by the user. Instead, \textit{behavior fidelity} was suggested, although this has the same limitations. Alternatively, \textit{body fidelity} was suggested. However, this has an overly passive framing and could be misleading by implying only to represent the user's body as an avatar and neglecting voice or neural input. None of the terms are perfect, so we chose \textit{action fidelity} as the most intuitive and flexible compromise.

The experts also criticized terms for which we could not find any better alternative. For instance, \textit{input fidelity} was challenged because undetected actions never enter the system and, thus, are not indeed system input.  We argue that the user's intent and the potential of capturing all actions are decisive for being included in \textit{input fidelity} as the user would expect a high-fidelity system to detect everything.
Lastly, we observed that \textit{display} and \textit{rendering fidelity} have strong connotations with visual screens for some people. In our sample, this was especially pronounced in the industry. In the instructional material, we emphasized that any modality can be rendered and displayed.

\subsubsection{Theme: Application Criticism}
With an example of their own previous work, all experts utilized the \model\ to analyze and reflect on a user study or system design. This enabled them to assess its applicability in a realistic scenario. 
Some of the experts experienced the onboarding as demanding: ``It's a lot to take in. [...] If I would have to use it as a tool, I would require a bit of time to get familiar with all the intricacies'' (E02). It was a challenge for the participants to understand the model's concept and remember the components' definitions. Some of them grasped the ideas quickly as they studied the subject intensely. To others, the proposed way to contemplate fidelity was unfamiliar, so they needed more time, detailed explanations, and examples to wrap their head around the model. Most participants asked several comprehension questions during or after the presentation of the model.

Even interviewees experienced in fidelity research recommended creating accessible instruction material: ``ideally, a graphical version that would make it very easy to use'' (E02). To enable teams to use \model\ as a practical tool, there must be a quick and easy way to learn how to work with it. Therefore, we developed an informative and intuitive poster that gives a visually appealing introduction to the model.  E02 suggested
``creating a very nice graphical representation of it that you can print out in A0 and have in your lab. [...] Having all of it up on a wall would be very good for me.'' This agrees with our experience of hanging it on the wall as a poster and referring to it during our daily work. 
We also prepared a slide deck that can be conveniently adapted for teaching material. Both can be accessed in the appendix and used for free under a Creative Commons license. 

All experts but two stated that they would like to use the model in their work in the future. One of the two, E08, wanted to read the paper first. The other expert, E10, was concerned about adequate validation before employing it: ``I think validation is important when it comes to something like this.'' E04 appreciated the expert interviews as a validation method: ``I like the approach. Very systematic.'' For E10, however, to sufficiently verify the soundness and acceptance of the proposed model, it must prove itself in practice on a large scale. Only if the community can work with it and embraces it the \model\ will be valid to E10. At the same time, this seemed to be the most promising prospect: ``At the moment, it's a theoretical model, a conceptual model. Making it more than that, that's a research opportunity. Bring it to practice!''

When applying the model, some experts fell into application traps, which led to criticism of the model's applicability. We collected these traps in Section~\ref{sec:traps} as guidance for readers without elaborating on them further here. We also collected special use cases and integrated them at appropriate parts of this publication, such as the applicability for multi-user systems (E06), mixed reality (E13, E06), or troubleshooting (E12, E14).

\subsubsection{Theme: Application Strategies}
From the observations in our interview workshops, we can conclude that the \model\ successfully guided the participants through the analysis process and was a helpful basis for discussion. This is evident in the different approaches and outcomes from the first and second tasks, i.e., without or with the help of the model. 
In the first task, without the model, the participants' explanations and considerations were unstructured, and occasionally seemed lost in the complexity of the subject. Most were uncertain about where to start and stopped after mentioning only a few aspects. 
In contrast, equipped with the model in the second task, all participants identified many more relevant aspects and kept elaborating on the interactions' fidelity. Typically, it took them a few moments to think about it and match the model with the use case. Then, they went through the single aspects and their connections, talking like a waterfall and coming up with additional insights they had not considered in the first task. Further, they now involved every part of the interaction in their analysis. E12 described this experience with: ``I thought it was really cool because [with the model] you could just work through several points step by step, whereas before I just groped in the dark.'' With this, the \model\ seemed to have an empowering effect on our participants. 

The experts used different strategies for applying the model to their individual use cases. We did not instruct them how to proceed but just set the task to describe interaction fidelity using the model. Some explicitly asked for the correct way to use it: ``Where does it start? On the top with the user? Or with action fidelity?'' (E12). 
Indeed, half of the sample adapted the approach we chose for presenting the model, starting with action fidelity and continuing in the loop aspect by aspect. As a variation, E02 started with action fidelity but then followed the logic of the interaction technique. Alternatively, three experts started with their study's independent variable and went in the loop from there. Similarly, E01 started with the aspect most important for the application but then went through aspects chaotically and went back and forth, addressing different modalities. E04 and E09 had a purely chaotic approach, following their narrative organically as it evolved. 
E14 deliberately made two rounds in the loop: first for a ``clean'' version of the interaction technique, then for the modified version with sensory manipulation. This helped him contrast the differences. 

It was our impression that the experts from the industry generally needed more information and assistance for applying the \model. For example, E12 and E07 asked to return to the introductory slides with the descriptions for performing the second task. As a result, their analysis was conspicuously close to the notes and talking points on the slides. Although it seemed more demanding, they appreciated how beneficial it was. E09 concluded that ``a more analytical approach for designing [immersive] interfaces and experiences would be reasonable because I think a lot of designers shoot in the dark.''

The experts in our sample used the model for different goals. In most cases, they roughly located all aspects within the fidelity spectrum as a starting point. After this analysis phase, they interpreted their findings depending on their goal. For example, some attempted to reason or predict experiential fidelity through the other aspects (E01, E05, E06, E09, E12). Others wanted to set informed design priorities by weighing the aspects regarding their relevance for a use case (E01, E06). For this, they also linked their interpretations to established design principles and effects, such as visual dominance.
For some researchers, assessing the degree of fidelity was less decisive. Instead, they wanted to identify which components the independent variables of a user study were part of (E03, E14): ``This is helpful! Because it makes more precise what I did and did not manipulate in my research'' (E03). This researcher realized that all of their manipulated study variables affect simulation fidelity, which they first criticized as not that interesting. Still, while contemplating using the \model, they realized they had achieved what they set out to do in the study: alter aspects of the interaction influencing plausibility illusion. Since E03 interpreted this as the relationship between simulation and experiential fidelity, they concluded to have chosen a suitable focus.
In a further step, some researchers reflected on aspects that would be interesting to manipulate in follow-up studies (E02, E06): ``It has been very useful trying to understand what it is I'm manipulating, and when designing an experiment, figuring out what are the factors that are relevant to manipulate'' (E02). 

A further reoccurring strategy of applying the model was identifying aspects of utmost importance for specific use cases or purposes (E06, E09, E13). For example, in applications for motor skill training, it is critical to optimize for high action fidelity, while simulation fidelity must be as high as possible for educational purposes. Therefore, some participants tried to define target variables, such as learning success, confidence in use, fun, or control precision, to deduce the determining components. 
Overall, the strategies and paths of applying the \model\ were as varied as the participants' objectives.

\subsubsection{Theme: Contributions}
Although the purpose of the interviews was to obtain criticism from experts in the field, we also want to outline the commendatory remarks of our participants. This theme includes positive remarks, suggested use cases, and contributions of the \model\ brought up by the experts. 

Overall, the interviewees appreciated the model: ``It's a nice theoretical framework. It's well polished'' (E04). Many of them enjoyed contemplating and discussing VR interactions with the model as the basis of the conversation: ``Interesting! That gives me some thoughts'' (E08). The model was described as ``super cool! super exciting! super useful!'' (E06), and E07 concluded that it ``makes sense! It's all comprehensible''. 
An interviewee from the industry liked the \model\ as it helped reflect on fidelity in VR: ``It's interesting because I understand it intuitively, but I don't know how to formalize it. And this is spelling it out for me. [...] I love it!'' (E09). 

E02 acknowledged that the model builds on existing frameworks and concepts: ``I like the model. I used similar models myself, but this one takes it a step further and adds detail. I think it makes really good sense. [...] It adds structure to a large body of literature. It distills a lot of different concepts and presents them in a single model. The fact that we can go one level deeper and get more sub-components, I think, has a lot of utility---both for designing studies and as a pedagogical tool.'' Contrasting it to similar frameworks, E11 liked the ``agnostic approach''. 
Extending previous work, E05 found that ``comprehensiveness and consistency around the entire loop of VR interactions is the main benefit.'' 
In particular, the advancements compared to the Framework for Interaction Fidelity Analysis (FIFA)~\cite{mcmahan2016.UncannyInteration} were addressed in the interviews. While the FIFA only considered input, the \model\ also considers the system output. Also, FIFA's focus on the system was extended to the user and their mutual influence. And while FIFA was limited to realism, the \model\ extends to any reference interaction, from reality or not. 

Further, it seemed important to many of the experts ``to have a precise and integrated set of terminology. I think that's good and the biggest benefit'' (E03). For unambiguous communication within the community, participants found it ``very useful to get a shared vocabulary'' (E02). Beyond using the same labels, E13 emphasized: ``The benefit is not only to talk about the same thing but also to trust the other person refer to the same thing.''
While we also received critical feedback and alternative suggestions, most participants liked the chosen wording. The terminology was described as ``accurate'' (E03) and ``consistent across the model'' (E02).

Most of the experts were eager to use the \model\ in their work, described it as ``helpful'' (E01), and found various use cases: ``I like this a lot in many ways!'' (E05).
Many of the proposed applications concerned system design and development. One utility often mentioned was the help in understanding the single components of interaction fidelity to predict experiential fidelity or the user experience. Distinguishing the fidelity aspects was also described as helpful in setting priorities depending on the goals or purpose of a system. As a next step, the model was considered helpful for iterating system designs to increase interaction fidelity. Also, participants liked comparing two similar systems based on the model. 

To analyze existing systems, the \model\ was used by some participants like a checklist: ``It's cool to think it through in small steps: like check boxes that you can tick off'' (E12). Also, E01 emphasized: ``The structure is helpful when going through the individual fidelity categories.'' Similarly, troubleshooting was mentioned as a use case. If there is an issue with the system or the users are unsatisfied, E12 and E14 suggested using the model to search for the problem systematically: ``I like how you can examine a user's experience in smaller steps: Where exactly now does the error get into the system?'' (E12). 
E06 and E14 further suggested use cases outside VR, such as for mixed reality interactions, video instructions, or other applications with a simulation approach.

Furthermore, the experts suggested numerous ways the model can be used for research. In the interviews, many researchers used it to understand a study better in hindsight: ``What is it that we manipulated in our study?'' (E02). We observed many instances where the experts tried to identify the modified independent variable of a study and how they searched for dependencies within the loop. This was also considered helpful for planning upcoming studies. 
Several researchers suggested a systematic literature review to identify which fidelity aspects were investigated in VR user studies. The model could provide a structure for a large body of literature (E02), and it could be expanded to be a signpost to all related work with details on specialized subcomponents (E13). Nine interviewees mentioned the significant research opportunity of systematically studying the fidelity components across the loop, both individually and regarding their influence on each other. Due to the model's structure, high comparability might help identify patterns (E05, E06, E13, E14), such as uncanny valleys in different sensory modalities (E04, E12). These endeavors were often considered a long-term community effort. 

Lastly, teaching was mentioned repeatedly as an ideal use case for the \model: ``Excellent presentation! I would include this in my curriculum right away. These are good learning materials for the students'' (E13). The participants liked how they could convey the processes and connections of different factors in VR interactions with one central figure as an overview.

\section{Application of the Model}
\label{sec:application}
In this section, we illustrate ways to use the \model\ in the work of researchers, designers, developers, practitioners, teachers, and students who work with VR. We collected best practices from our own experience and the expert interviews (see Section~\ref{sec:validation}), possibilities for applying the model as a tool, and common traps to be careful of. 


\subsection{Best Practices}
\label{sec:bestpractices}
We recommend the following guidelines for applying the \model{} purposefully and effectively. They build on experiences from using the model for over two years and observing the participants in our expert interviews use it. 
\begin{itemize}
    \item \textbf{Mind your reference.}
It is crucial to keep the reference interaction in mind when assessing fidelity. Define the reference as specific and detailed as possible before comparing it to the VR interaction. Keep recalling the reference interaction throughout the analysis, not only at the start. When using several references simultaneously, be aware of which one you currently compare to. 
    \item  \textbf{Choose a focus.}
Decide how holistic or focused your analysis should be. Are you interested in just one isolated aspect of the experience (e.g., a handshake) or all elements involved (e.g., the full complexity of greeting conventions, environmental conditions, multiple users, etc.)? It can be easier to break down the interactions for analysis. 
    \item  \textbf{Set a goal.}
Reflect on the purpose of applying the model. The \model\ can be used as various lenses through which interactions can be examined, allowing different perspectives. We propose a number of ways to apply the model in the next section.
    \item  \textbf{Skip irrelevant aspects.}
You can use the \model\ modularly. In most cases, only parts of the loop are needed for analyzing an interaction. Feel free to ignore irrelevant components and instead focus on the key aspects. 
    \item \textbf{Be objective.}
The connotation of ``high-fidelity'' as ``better'' is common in practice as the \textit{Better--Worse Trap} in Section~\ref{sec:traps} illustrates. However, the \model\ works best as an objective tool. The fidelity concept is free of judgment. Therefore, clearly differentiate between ``high experiential fidelity'' and ``great user experience.'' While higher fidelity can be desirable for interactions, deliberately decreasing fidelity can also help reach a goal. 
\item \textbf{Justify your assessment.} 
Be more specific than only assigning low- or high-fidelity labels. The insights from analyses or discussions can be richer if you argue how you came to that conclusion. This can help identify dependencies or patterns.
\item \textbf{Adhere to the terminology.} 
Please stick to the official terms used in the \model\ when referring to it in scientific communication. If you need to specify a fidelity subcomponent (e.g., haptic fidelity), additionally mention the higher-level component it belongs to in the model (in this example, display fidelity). 
\end{itemize}

\subsection{Lenses of the Model}
The model can serve as different lenses through which interactions can be viewed. Depending on your goals, the \model\ can be applied with various strategies yielding different insights. 

\textbf{Describe \& Report}: The \model\ provides well-founded and consistent terminology for the different aspects of fidelity in VEs that all contribute to the overall fidelity of a system. This structure helps to identify and name the fidelity aspect of interest and provides a foundation for an informed and differentiated discussion about fidelity in VR. In communicating study results, the investigated realism dimension can be unambiguously specified, supporting comprehension and retrieval by other researchers. Furthermore, the model offers a structure for reporting VR system specifications rigorously within the VR research community.

\textbf{Understand \& Distinguish}: The \model\ can be used to understand the influencing factors on the overall fidelity in VEs at different stages of the HCI loop. It facilitates a comprehensive understanding of distinct factors and typical relations between the single components that contribute to the overall fidelity in VR. The model can also help to untangle interwoven components such as input and output through the same device. For example, user actions and haptic feedback when using force-feedback gloves directly depend on each other. 
Designers and developers can apply the framework to understand why their product performs the way it does and make informed decisions that explicitly address the influences of the single fidelity components on the overall fidelity. This could, for example, include a coherent level of fidelity across all stages of the loop or strategies to compensate for limitations in one component through improvements in others, e.g., when bodily trembling in an interaction can be compensated through denoising the signal with the transfer function. Overall, educators can use the model to explain the complexity of fidelity in VR and give students an overview of approaches for the different components of fidelity and examples of how these typically unfold in combination.

\textbf{Compare \& Analyze}: Researchers, designers, and developers can apply the \model\ to compare variants in VR setups, e.g., different input devices, and systematically analyze the effects on the distinct fidelity components. As the model holistically addresses input fidelity and output fidelity, a complete reflection of the effect of, e.g., joystick vs. actual walking as different input techniques for locomotion in VR not only addresses the fidelity of the users' actions, the detection and the transfer function on the input side. It also directs to connected aspects of the output, such as perceptual fidelity, which can broadly vary between setups with different input fidelity.
The model's components can provide insight into where differences or similarities between systems are, how decisive they are, what the underlying reasons are, or what could be done to compensate for them.

\textbf{Hypothesize \& Guide}: The \model\ can be used to generate research questions around fidelity in VR to analyze and understand this design space further. It reveals many research opportunities, such as design guidelines regarding the combination of different fidelity levels in the model's components. For example, \textit{can components compensate for each other? Or should the components rather have coherent levels of fidelity? How do the quantitative components influence the overall experiential fidelity?} The model could also be used as a starting point for heuristic evaluations of the fidelity of VR systems by systematically addressing the fidelity components. For empirical evaluations of realism in VR, researchers can also use the \model\ to formulate reasoned hypotheses as well as to explain and discuss the findings in relation to the different stages of the loop. In the following section, we will elaborate on the arising research opportunities based on our model.

\textbf{Teach \& Convince } 
The model's simple structure combined with the intelligible visualization makes it easy for students to learn about VR interactions and how different components must be considered for reproducing something virtually. As the \model\ can be universally employed for any use case, it works in various practical and scientific curricula. Teachers can use the sequence of aspects to guide students stage-wise through the relevant components of interactions while emphasizing the reciprocal nature of human-centered simulations.
Similarly, the model is suitable for convincing stakeholders and managers with limited experience with HCI methods or VR technology why a proposed strategy, system design, or research agenda would be advisable. The complexity of seemingly trivial interactions can be demonstrated just as effectively as the interdependence of the single components.

\subsection{Application Traps}
\label{sec:traps}
We identified several common pitfalls that people applying the model fall into and which we were also repeatedly caught in when developing it. The following traps are partially based on experiences from the expert interviews presented in Section~\ref{sec:validation}. Be sure to avoid these common mistakes to get the most out of the \model.

\textbf{Better--Worse Trap} 
The most frequent fallacy might be attributing a judgment instead of objectively describing fidelity. People tend to use phrases such as better, worse, nicer, more immersive, better UX, etc., instead of an impartial assessment of the exactness of correspondence. The neutral, dispassionate view in fidelity evaluations is not the most intuitive attitude. A possible explanation is that fidelity and desirability correlate for many interactions, especially when striving for natural interfaces. However, it can also be beneficial to decrease fidelity deliberately to achieve a particular effect. Therefore, it helps to avoid thinking of low fidelity as a shortcoming.

\textbf{Time-Travel Trap} 
In contrast to absolute assessments of how close an aspect is compared to the reference, peoples' assessment is sometimes linked to the state of the art at a certain time. For example, some researchers evaluated a system in the context of technical possibilities at the time of development. Consequently, the low-poly visuals of an application were described as high-fidelity because 15 years ago, when it was built, the system was considered world-class, thus high-fidelity compared to anything else at the time. 
It can be reasonable to make a time-dependent comparison, e.g., when selecting hardware with the current technical limitations or tracing system capabilities over time. In other instances, however, it can be misleading to make assessments depending on the current state of the art as it shifts the assumed upper limit from \textit{maximum fidelity} to \textit{currently attainable fidelity}.

\textbf{Apples-and-Oranges Trap} 
Another trap we experienced and observed frequently is comparing the VR interactions to different references without noticing. For example, in the expert interviews, one participant assessed a locomotion technique as low- and high-fidelity at the same time and was confused about the outcome. The reason was that the comparison was made to different references: first to a teleportation technique, second to actual walking. While both comparisons can result in valuable insights, it should always be clear which reference is currently compared to.
Another person fell into the same trap and referred to it as ``comparing apples and oranges.'' Therefore, it is vital to apply the \model\ in a meaningful way. When comparing, choose a clear purpose and an adequate reference interaction. When in doubt, you can ask: \textit{What} do I want to reproduce?

\textbf{Visual-Dominance Trap} 
As there is a strong connotation of \textit{rendering} and \textit{display} regarding vision, there is a risk of neglecting other modalities and senses when using the model. In many experiences, the visual impression is the most dominant and sophisticated modality. But it might be important to include all senses, depending on the use case. On the other hand, it can also make sense to narrow the focus to relevant modalities. 

\textbf{Feasibility Trap} 
We also observed the risk of overestimating a system's fidelity and referring to it incorrectly as \textit{maximum fidelity}, although it was merely the highest fidelity possible to achieve. Increasing it further could be restricted by limited resources, physical boundaries, personal abilities, or one's own imagination. 
However, it is irrelevant to assessing fidelity if there are limits in feasibility. Describing the exactness of correspondence is technology-agnostic, hence independent of the reasons behind a system design. To achieve maximum fidelity, optimizing a system as much as possible is insufficient unless there is a perfect match with the reference. We may never accomplish maximum fidelity in some aspects.

\section{Discussion}
\label{sec:discussion}
We now turn to the broader context of the model, considering its general meaning and implications. In the following, we discuss potential patterns of how the fidelity components might be connected, considerations of optimizing for realism, how the \model\ relates to similar constructs, how fidelity can be described objectively, how it can be measured, the significance of perceived fidelity, and how the model can be applied to reference frames other than reality, such as fiction or mixed reality.

\subsection{Patterns}
\label{sec:patterns}
Investigating numerous VR interaction techniques, systems, experiences, and user studies, we found similar phenomena repeatedly. From this, we distilled reoccurring connections, dependencies, and relationships between the fidelity components. In this section, we propose potential patterns that might be discoverable in various VR interfaces. Not all patterns necessarily appear in every interaction. 
Please note that the patterns presented here are not systematically studied and lack empirical evidence. They are merely based on incomplete sets of examples and theoretical reasoning. Further systematic research is needed to test these suggested patterns and reveal additional ones. 

\textbf{Bottleneck Pattern |} 
In some use cases, experiential fidelity cannot be higher than a limiting key component, even if other aspects have much higher fidelity. For example, when juggling in VR, it is irrelevant if visual rendering fidelity is exceptionally high as long as action or simulation fidelity is low. The bottleneck of these limitations will always impair experiential fidelity. 
Consequently, it is necessary to identify the critical components of an application and prioritize them for increasing fidelity. 

\textbf{Loss-Propagation Pattern |} 
Certain aspects in the loop cannot be higher than the previous one. In this case, the component inherits the constraints of the preceding one. We can find such a conditional dependence in the pipeline \textit{Simulation}~→ \textit{Rendering}~→ \textit{Display}. For example, a display can depict something at most with the resolution it was rendered at, and the rendering software can, at best, match the quality of the simulation's 3D model. Although we can speculate and approximate to compensate, we cannot assure the compensation's authenticity.
Thus, loss at an early stage of such sequences cannot reliably be made up for and is propagated throughout the loop. 

Similar dependencies can be in the sequence \textit{Action}~→ \textit{Detection}~→ \textit{Transfer}. For example, if a controller prevents a natural hand pose when juggling, no sensor of an input device can make up for this, and if the sensors do not detect the hand position, a transfer function cannot compensate for missing data. 
There might be further such sequences, e.g., \textit{Display}~→ \textit{Perceptual}~→ \textit{Experiential}; or \textit{Experiential}~→ \textit{Action}. 
As a consequence, it is advisable to optimize at the start of such sequences to avoid inheritance of early losses. 
However, a component can also be limited by a component much earlier in the loop, e.g., \textit{Action} → \textit{Perceptual} due to the vestibular system. Therefore, searching for the root of an issue in the preceding components can be helpful, as it might just be a propagated problem. 

\textbf{Irrevocable-Loss Pattern |} 
An extreme version of the Loss-Propagation Pattern was proposed by E08 in our expert interviews: ``It's like a pipeline. [Progressing through the pipeline], there are only losses.'' 
In many use cases, such a drastic error progression might occur. However, it can be prevented in some cases. For example, if the hand pose is detected incompletely due to occlusion, we can still compensate for this deficiency by reconstructing a probable hand pose using anatomic models and inverse kinematics.
Consequently, there might be a viable solution to a component's limitation later in the pipeline, compensating for earlier losses. 

\textbf{Free-Upgrade Pattern |} 
Maximum fidelity in some components automatically leads to maximum fidelity for the subsequent component. For example, if the displays reproduce the physical stimuli perfectly, they will also be perceived indistinguishably from the original stimuli of the reference, as we cannot sense the source of a physical signal but just the signal itself. Thus, maximum display fidelity inevitably results in maximum perceptual fidelity. 
Similarly, experiential fidelity will automatically be at maximum as inherited from the previous component. 
Therefore, if you need to increase the fidelity of a particular component, it might be helpful to optimize the preceding components.

\textbf{Uncanny-Valley Pattern |} 
Comparing varying input fidelity with user performance, \citet{mcmahan2016.UncannyInteration} suspected an ``uncanny valley of VR interactions'' that leads to poor performance for unfamiliar interfaces with medium input fidelity. While the tested low-fidelity interfaces were known from preexisting systems and the high-fidelity interfaces were intuitive, both delivered higher performance than unknown, somewhat abstract medium-fidelity alternatives. However, as the authors discuss, this non-linear relation cannot be generally applied to any input system. Further studies identified similar patterns in gaming~\cite{lukosch2019.FidelityGaming}, training~\cite{bhargava2018.IntFiContinuum}, and locomotion~\cite{nabiyouni2017.DissLocomotionFidelity}. 
Consequently, it might be beneficial to avoid interaction techniques with medium fidelity.

\subsection{Optimizing Realism}
Reality is not the only reference frame to which the \model\ helps compare VR, but it is undoubtedly a particularly relevant and common one. Therefore, we discuss the ambivalent goal of striving for high realism. 
An interaction's level of realism is merely a descriptive, impartial attribute. High realism is not \textit{per se} superior. For some use cases, increasing interaction realism can be beneficial (e.g., skill training and education) or fundamental (e.g., preservation of historical artifacts). For other use cases, enhancing realism might be negligible (e.g., data visualization, art), detrimental (e.g., fictional entertainment), or even harmful (e.g., source memory confusion). Supernatural abilities have massive potential for VR interactions (e.g., brain-computer interfaces for telekinesis or changing laws of physics), as pointed out by \citet{bowman2012.Naturalism} and supported through a design method by \citet{sadeghian2021.superpowers}. 
Even in use cases originally meant for reproducing real experiences, such as social interactions, deviating from high realism can enrich the experience and introduce new possibilities~\cite{mcveigh-schultz2021.WeirdSocialSuperpowers,mcveigh-schultz2022.BeyondVRmeetings, bonfert2023.VRmeetings}. 
\citet{dewitz2023.MagicInteractionsFramework} develop a framework on interaction techniques beyond realism, such as magic techniques, superpowers, or hyper-natural augmentation. It locates interaction techniques along the three orthogonal axes internalizability, congruence, and enhancement. 

Striving for high fidelity requires time and financial effort. Therefore, it is important in research and development to critically reflect on how much realism is desirable and expedient. The \model\ can help identify components that should be optimized or can be less prioritized. We discuss such considerations in Subsection~\ref{ssec:optimization}.
While we can do a lot of good with highly realistic VR simulations, it can also be harmful and used maliciously. A growing body of literature addresses problematic implications of progressing XR technology, the ethics of increasingly attainable realism, and the risk of hostile manipulations~\cite{slater2020.EthicsXR, madary2016.EthicalConductVR, wassom2015.ARethics, tseng2022.DarkSideVR, rubo2021.SourceMemory}, which should be considered when striving for high-fidelity applications.

Supporting the recommendation of careful tradeoffs by \citet{jacob2008reality} as part of their framework for reality-based interaction, we further suggest reducing realism in return for other desired qualities that align with the simulation's purpose. \citet{jacob2008reality} propose considering benefits in expressive power, efficiency, versatility, ergonomics, accessibility, or practicality for a tradeoff.
For example, in a training simulator for learning how to juggle, high action and simulation fidelity are essential for the trainee to transfer the acquired skills to reality and apply the movements with real balls. Nevertheless, offering a training mode in slow motion to practice the movements without time pressure can be a beneficial deviation from reality.

\subsection{Related Constructs}
\label{sec:DisRelatedConstructs}
In the literature, numerous concepts and ideas have been associated with the fidelity of a simulation, such as the Place Illusion of ``being there'' and the Plausibility Illusion (also referred to as presence)~\cite{slater2022.UpdatePresence, skarbez2017.Presence}, coherence\cite{skarbez2016.DissertationPsi}, immersion~\cite{bowman2007.EnoughImmersion}, engagement~\cite{lukosch2019.FidelityGaming}, and others~\cite{skarbez2017.Presence}.
We consider these as different from but correlated with fidelity. Thus, a high-fidelity interaction could result in low presence but usually leads to high presence. Conversely, a high sense of presence in a coherent, highly engaging world can also be achieved with a low-fidelity system. 
\citet{slater2009.RealisticResponses} argues that high levels of place and plausibility illusions lead to realistic behavior of the user. Accordingly, the user's reactions to the virtual experience ought to correspond to how the user would react to the reference interaction, i.e., high experiential and action fidelity. 

In practice, these constructs have a strong link and correlate in countless empirical studies. They also overlap in their typical assessment, e.g., some presence questionnaires comprise items to self-report perceived or experienced realism~\cite{witmersinger1998.PQ,schubert2001.IPQ}. Yet, the constructs concern different theoretical questions. It is, therefore, important not to confuse their claim. In particular, in empirical evaluations, the choice of measurements and interpretation of evidence depends on the concept that the research question revolves around. 
While there have been decades of discourse on the conception of presence and similar concepts~\cite{skarbez2017.Presence}, fidelity as the objective degree of correspondence between simulation and original is straightforward and with the distinct components of the \model\ intuitive for the planning and analysis of VR systems. Although this makes fidelity an unequivocally defined concept, it might not be the relevant one to evaluate depending on the purpose of a system or study, just as presence is not always the essential metric that should be sought after~\cite{latoschik2022.CongruencePlausibility}.

\subsection{Describing Fidelity Objectively}
\label{ssec:quali}
Let's assume there is an objective, indisputable ground truth of how exact the correspondence between an original and its replication is. This truth could be described objectively if it is known. The assessments of what is true, however, can be subjective and might diverge. The more precisely we agree on how to evaluate interaction fidelity in a systematic, replicable way, the more objectively we can determine and agree on it. As discussed in the following subsection, we can only approximate the ground truth and achieve consensus through standardized measuring criteria.
Technical parameters regarding system fidelity are simpler to assess objectively, while we need to rely more on subjective evaluation of user-related aspects. As various aspects determine the multi-faceted concept of interaction fidelity, it is difficult to identify its ground truth comprehensively. 

Here is an example of a seemingly unambiguous and objective fidelity assessment. Probably, most people would agree that compared to the reference interaction of grasping an object, we can attribute higher interaction fidelity when the user reaches out and encloses the virtual object with their bare hand, than when the user points at the object with a hand controller and presses the trigger button. The latter implementation relies on mappings and seems less natural.
However, primarily action fidelity is higher in the first implementation, while display fidelity is lower due to the lack of haptic feedback. The controller's passive force feedback has a higher correspondence to grasping a rigid object, providing higher display fidelity. 
Hence, depending on the focus or context, evaluations can vary. Although there is an irrefutable ground truth that we strive to ascertain, we do not necessarily succeed in doing so objectively. As a result, it is essential in communication---especially in scientific discourse---to clearly describe our perspective and reasoning. 

Especially the \model's components Perceptual and Experiential Fidelity are difficult to determine objectively as the subjective nature of a person's qualia is individual and might be intrinsically subjective. Depending on the perspective on the philosophical mind--body problem, it might not even be possible to deduce mental events of the unobservable mind from the physical events in the observable brain~\cite{georgiev2020QuantumMindBrain}. Therefore, an objective, holistic description of interaction fidelity might be unattainable.
Regardless of metaphysics and seen from a pragmatic point of view, it is more expedient to assess experiential fidelity subjectively on a user-by-user basis: Does a specific person experience the VR simulation exactly how this person would experience the reference interaction? This perspective takes all personal characteristics and biological features into account. Usually, the diversity of people's individuality must be considered because most systems are aimed at large user groups.

\subsection{Measuring Fidelity}
\label{ssec:quanti}
Beyond a qualitative understanding of interaction fidelity, it can be helpful to express the degree of correspondence quantitatively. 
Some fidelity aspects can already be assessed with high objectivity. For example, characteristics of input and output devices can be technically gauged, e.g., regarding pixel density, sensory noise, or degrees of freedom. Some of these quantifiable parameters allow a direct interpretation of how they affect interaction fidelity in direct comparisons. For example, a screen with a higher resolution than an otherwise identical screen provides higher display fidelity. For other aspects, it is harder to infer an uncontroversial effect on interaction fidelity from technical parameters. For example, regarding rendering fidelity, the influence of shaders treating light reflections differently might depend on various circumstances and is more intricate to interpret. 

Some aspects are commonly assessed subjectively with self-reports, such as in questionnaires or interviews. Due to their subjective nature, experiential and perceptual fidelity are usually measured through user reports. But also some system-related components might only make sense to be assessed subjectively by large numbers of evaluators, for example, the credibility and human likeness of virtual human animations. 
Unfortunately, there is a limit to how much we can ask users to share their pmpressions in studies, making a holistic assessment of all subjective parameters impossible. However, we claim that it is possible to predict experiential fidelity sufficiently if enough about the other fidelity components is known.

In this work, we described the level of fidelity with the coarse categories low, medium, high, and maximum fidelity. This gives us an approximate location on the continuum and allows the rough comparison of a few systems. However, we should strive as a community for detailed, theory-based, technology-agnostic, and unambiguous metrics for all fidelity components, as outlined in the research opportunities in Section~\ref{sec:researchOps}.

\subsection{The Normative Power of Subjective Truth}
\label{ssec:normativepower}
There is an additional challenge when comparing VR interactions to the real world. In the case of assessing realism, we need to agree not only on the nature of the simulated interaction but also on the reality-based interaction.
From a philosophical perspective, it is hard enough to agree on what ``reality'' objectively is. Anybody discussing the manifestation of the real world can only do so from their subjective point of view informed by their individual perceptions. While we can assess the exactness of the correspondence between a simulation and what is considered a broad consensus about the real world, the judge will ultimately be the users with their impressions. Depending on the simulation's purpose, their subjective judgment may not be decisive for how the interactions are designed, but often, experiential fidelity is the only outcome that matters and will be optimized for. 
In this case, only the user-related components regarding what the user perceives, experiences, and does seem important. Why should we then care for system fidelity at all? The system-related components primarily determine the levels of the user fidelity aspects. For designers and developers of VR systems and interactions, system fidelity is the only way to influence the user's perception, experience, and actions. The better we understand the components' mutual influence and interdependencies, the more effective our endeavors can be.

Interestingly, increasing the fidelity of single aspects does not necessarily increase the experiential fidelity. Previous research has suggested that reducing fidelity aspects for some interactions elicits higher experiential fidelity~\cite{nilsson_DecreasedFidelity_2017,mcmahan2016.UncannyInteration}. For example, when moving through a VE using a treadmill as an input device, the walking speed is experienced as more realistic by the user when the virtual pace is exaggerated relative to the originally slower pace in the real world~\cite{banton2005.SpeedPerception,kassler2010.TreadmillSpeed,powell2011.VisualSpeedGain}. Here, the transfer function is mapped unrealistically to compensate for other limitations in interaction fidelity, such as the missing kinetic feedback from staying in place (i.e., low perceptual fidelity). The required amount of exaggeration has also been found to depend on the visual display fidelity: The smaller the field of view is, the stronger the speed must be exaggerated for the user to feel realistic~\cite{nilsson2014.NaturalSpeed}. Alternatively, the visual projection can be distorted to display more peripheral information~\cite{nilsson2015.DisplaySpeed}. Hence, reducing transfer or rendering fidelity can increase experiential fidelity. 
Similarly, \citet{bowman2012.Naturalism} argue that high fidelity in a certain component (e.g., action fidelity by rotating a Wii controller to steer a vehicle) might result in lower perceived realism because of the shortcomings in other components (in this example, missing force feedback and latency). 
As a consequence, the purpose of a system must be considered for prioritizing the different components' targeted level of fidelity.


\subsection{Applicability to Fiction, Mixed Reality, and Other Reference Frames}
The \model\ is designed to help assess the correspondence of interactions in virtual reality with interactions in any other reference frame. While the reference can be the real world, the model can also guide the analysis of fictional and other scenarios as long as the element to be reproduced is explicitly specified. This applies to any fictional media, imaginary narrative, dream, or fantasy, but also other VR systems, setups of previous user studies, or different levels of blended realities along the reality--virtuality continuum~\cite{skarbez2021.RVcontinuum}. 

As an example of fictional references, the characteristics, behavior, and appearance of a lightsaber from \textit{Star Wars} are extensively defined by the creators of the fictional artifact and, therefore, can be virtually reproduced. By contrasting the simulation with the original descriptions and depictions, we can evaluate its fidelity, as demonstrated in Example 3 of Section~\ref{sec:examples}. The less clear the original to be simulated is, the more controversial the fidelity assessment might be. To stick to the Star Wars example, when simulating interactions using ``the force'' for telekinesis, it is less obvious how this might be realized virtually---arguably not only with a hand gesture but rather a brain-computer interface. The experiential fidelity will broadly differ between users as they have differing conceptualizations and expectations depending on what Star Wars media they have seen or read. 

While developing the \model, we encountered scenarios in which no reference would be obvious to compare to, which always led us to the question: What are we trying to (re)create here? As an example from our expert interviews, participant E09 brought up an application to teach chemical processes. However, the equivalent process from reality on an atomic level made no sense to reproduce virtually for teaching. Instead, it had to be magnified to a human scale. But what could it sound like if two atoms bond on a human scale? The most suitable reference interaction we could come up with was the educator's idea of what an upscaled version of the virtualized school book model might look, sound, and feel like. This is where the model reaches its limits. A comparison using the \model\ provides little insight if the reference is only vaguely defined. 

Another use case for applying the model is to compare two interactive systems. For example, we could compare the realism of playing baseball on a \textit{Nintendo Wii} with an implementation in VR. While aspects of input fidelity might be similar using 6-DoF controllers, the VR version might show high fidelity in other components and explain outcomes of comparative user studies. 

Similarly, the \model\ can be used to evaluate mixed reality (MR) applications blending virtual worlds with physical reality. There are many forms of incorporating more or less portions of different realities~\cite{skarbez2021.RVcontinuum}. The interaction can be based in the real world with virtual elements integrated, or VR can be the foundation comprising elements from reality---any combination is conceivable. To apply the model meaningfully in MR contexts, it is even more important to clarify what is being compared. We advise treating the blended realities as unity and comparing it to a non-mixed equivalent. Consider, for example, an MR meeting in which co-located users participate in the real world and virtual users join remotely. To design the interactions in this blended setting, comparing them jointly to the purely real or a purely virtual equivalent can be helpful. 
Obviously, you can achieve high levels of fidelity when simply augmenting reality with virtual elements compared to the challenge of building a system that recreates everything virtually from scratch. Automatically, some aspects are maximum fidelity as they equal the real world. But as \citet{lindeman2009.MemoriesExpFid} proposed, why should we not leverage parts of the real world and augment it to create overall high-fidelity interactions if reality affords it, such as using passive haptics for teleoperation? 
On the other hand, we encounter limitations that are more difficult to resolve in an MR context. For example, a remote user cannot manipulate a physical object. Because the object is integrated into the interaction but not necessarily part of the system, maximum fidelity cannot be reached without virtually modifying the real world. 

Another special case we would like to address is multi-user systems. Here, we find another added complexity when comparing interactions because the same encounter or activity might be experienced differently. We recommend splitting every comparison per person to avoid entangling the actions and perceptions of users or the differing hardware available to the users. Each user's interaction with the system must be considered individually for insightful analysis. This is especially important in asymmetric settings, such as in MR, where users have different possibilities and restrictions in perceiving and influencing the simulation.

\subsection{Limitations}
Inductively built on established HCI theory, no empirical evidence confirms the structure of the \model. The model was reviewed, practically tested, and critically discussed in interviews with 14 experts from the field, but it has not been systematically evaluated with large numbers of users in the wild. The most conclusive validation will be the community's application of the model in everyday research and development, which is yet to be seen. 
To provide a universal structure, our model deliberately does not include media-specific fidelity focuses such as \textit{narrative realism} as described by \citet{Rogers2022.suchwow} or fidelity addressing the single senses (e.g., olfactory fidelity) but rather provides a generic framework, in which all of these can be further detailed. Depending on the use case, specific aspects can influence more than one component, such as visual fidelity, which can affect all three components of output fidelity. 
Another current limitation concerns the quantification of fidelity, which would help assess and compare approaches. We currently apply the approximate ranges of low, medium, high, and maximum fidelity similar to previous research~\cite{mcmahan2016.UncannyInteration, gilbert2016.Authenticity,lukosch2019.FidelityGaming} and we regard numeric assessments with standardized metrics as an opportunity for future research. 

The \model\ is designed for examining interactions in the context of VR. This does not necessarily involve a graphical 3D environment, multimodal interfaces, a head-mounted display, or a self-representation of the user. 
The model can be helpful in better understanding other human-computer interactions or even non-computer-assisted technology that involves some simulation, e.g., ship navigation simulators or telemedicine interfaces. 
However, we emphasize that not all definitions and concepts will fit perfectly. 
We encourage using the \model\ wherever it can provide structure and guidance but advise awareness of blurred lines of systems and realities that make identifying correspondences difficult.

\clearpage

\section{Research Opportunities}
\label{sec:researchOps}
Understanding the intricacies of building systems closely resembling the real world or other reference frames requires systematic research. The conceptual nature of the \model\ opens up foundational research avenues as it considers the full scope of people's interactions with virtual worlds and illustrates relations of the underlying processes. While it can be interesting to examine a component individually (e.g., action fidelity of a controller-based system), it inevitably depends on others (e.g., on display fidelity due to contact forces from holding the controller). The complexity of interdependencies, impacts on other components, and methodological challenges raise novel research questions for future work. Further, the model can inspire new perspectives on optimization and methodology.

\subsection{Relationships}

\paragraph{Which components depend on others?} 
On the system side, there are close relations between the hardware components and the software transmitting between them, e.g., for an output device, the rendering software calculating the simulation output for a display to show must be precisely matched with both attached components. The same applies to the transfer functions translating between input devices and the computer. But also, on the user side, we find a close connection between the sensory information from the receptors and the brain processing and interpreting it. Although it is helpful for scientific discourse and system development to abstract the individual components, they cannot be regarded as independent. 
Even components that are not consecutive in the loop show dependencies. For instance, when assessing a system with eye tracking, the action fidelity of a user's gaze inevitably depends on rendering and display fidelity. It can only be planned or evaluated together.
But how do the components generally depend on and influence each other?

\paragraph{Can one component compensate for another?} 
If one aspect of realism is constrained, can an increase of another fidelity aspect make up for it? If so, can any other aspect or only a specific one? For example, in a system without a haptic display for rendering forces, modifying the control/display ratio can still induce a sense of kinesthetic forces~\cite{rietzler2018.KinestheticFeedback, Samad2019.PseudoWeight}. As a consequence, there is higher perceptual fidelity despite low display fidelity by deliberately lowering transfer fidelity. Are there similar compensations that allow us to build cost-effective and universal systems? 
Another example was given in subsection~\ref{ssec:normativepower} concerning the exaggerated virtual walking speed when using a treadmill. Due to the lack of kinetic cues from not moving forward, the limitations in perceptual fidelity can be compensated by reducing transfer fidelity with a higher speed gain~\cite{nilsson_DecreasedFidelity_2017}. 
Similarly, if the head-mounted display's field of view is small, hence low display fidelity, a decrease in rendering fidelity due to minifying the visual output can result in higher experiential fidelity~\cite{steinicke2011.NaturalProjections}. Further, when using redirected walking as a locomotion technique, action fidelity is reduced as the virtual and real paths do not match due to physical restrictions in space. We can compensate by adjusting transfer fidelity~\cite{razzaque2001.RedirectedWalking}, simulation fidelity~\cite{suma2012.ImpossibleSpaces}, display fidelity~\cite{auda2019.EMSredirWalk}, and rendering fidelity~\cite{langbehn2018.BlinkingRedirWalk} all to maintain perceptual fidelity.
\citet{nilsson_DecreasedFidelity_2017} suggest ``that when limitations to a given component of fidelity reduce or distort perceptual information, then sometimes it may be possible to compensate by adjusting another component of fidelity---even if the adjustment on the surface constitutes decrease in the fidelity of the second component.''
Future research might investigate what other compensations should be considered in systems with restricted fidelity.

\paragraph{What components constitute experiential fidelity?} 
Every component ultimately influences experiential fidelity. Some components are already well-understood through years of research, such as the aspects of input fidelity. Other components still need more scholarly attention. Above all, it requires further research to understand how strong the components' impact on experiential fidelity is. 
We hypothesize that the influence varies between components and cannot be reduced to a simple weighted sum of the single components. Various neural phenomena will increase complexity, such as superadditivity in multisensory integration, i.e., the effect that, for example, visual and auditory stimuli give a stronger sensory impression combined than when just adding up the individual impressions~\cite{stanford2017._superadditivity}. 
It is still unclear whether the model's components are sufficient predictors for experiential fidelity. Other influences not represented in the \model\ might affect perceived fidelity. For example, \citet{witmersinger1998.PQ} integrate the meaningfulness of the experience in their questionnaire factor \textit{realism}. Consequently, a dilemma would arise about how a meaningless experience from reality would be effectively simulated.

The interdisciplinary nature of this component calls for joint research, especially including psychology and cognitive science. Unraveling how experiential fidelity relates, depends, and affects related constructs, such as presence, coherence, or user experience research, is a complex endeavor that has already been embraced~\cite{skarbez2017.Presence}, but must be pursued in further detail---both theoretically and empirically. 
We suggest systematic analyses with a study design similar to the experiment by \citet{skarbez2017.PlausibilityIllusion} on the influence of components of the plausibility illusion.

\paragraph{Which further patterns can be identified?}
Beyond the potential patterns that we proposed in Subsection~\ref{sec:patterns}, we can seek further patterns by systematically analyzing and linking empirical evidence based on the \model's structure. Similarly, most of our proposed patterns need empirical validation.

\subsection{Optimization}
\label{ssec:optimization}

\paragraph{What benefits result from improving each fidelity component?}
While it seems safe to assume that maximum fidelity has advantages for various user experience metrics, several studies demonstrate how less-than-perfect fidelity systems have considerable limitations in performance and preference~\cite{bowman2012.Naturalism}. Since it is an immensely long way to the ``ultimate display''~\cite{Sutherland1965.UltimateDisplay} indistinguishable from the real world, we currently ought to focus our research efforts on interactions with medium to high fidelity. Striving for maximum fidelity is costly and must be justified.
As outlined in the Uncanny-Valley Pattern in Subsection~\ref{sec:patterns}, medium-fidelity interfaces can even result in a worse outcome than a low-fidelity implementation.
\citet{bowman2012.Naturalism} hypothesized that hyper-natural interaction techniques (or ``magic'' interactions) could potentially even exceed the best possible performance of natural approaches. Yet, we need to recognize where to deviate from faithful imitation profitably. 

After decades of empirical research on VR interactions, a large body of literature informs us about the effects of different levels of fidelity~\cite{Rogers2022.suchwow, bowman2012.Naturalism, Gisbergen2019.RealismExpBehav, ragan2015.VisualComplexity, Grant2019.AnalysisOfRealism, bhargava2018.IntFiContinuum, lukosch2019.FidelityGaming}. Unfortunately, it is often not further differentiated what fidelity component varies or how we can relate the findings in a broader picture. With the model's systematic and theoretical foundation to understand interaction fidelity as an overarching concept, we can better connect the dots and systematically design follow-up studies.

\paragraph{What are the natural limits for expedient system optimization?}
Technically, we could indefinitely increase interaction fidelity by reproducing reference interactions in ever greater detail, converging closer and closer to maximum fidelity. However, human perceptual sensitivity and body control are naturally limited, making further optimization futile. For example, although screen resolution could be increased to the point where we can display a grain of sand at the horizon, no user could ever tell the difference as the retina's resolution is limited. Similarly, the limited precision of performing manual tasks makes more precise tracking obsolete.
Studies on such constraints in perception or body control can inform expedient system design when systematically assessed, e.g., as has been demonstrated for thresholds of redirected walking~\cite{langbehn2018.BlinkingRedirWalk, steinicke2010.RedirectedThresholds}, virtual hand offset~\cite{benda2020.HandOffset,zenner2019.HandRedirectionThresholds}, latency for foveated rendering~\cite{albert2017.LatencyFoveatedR}, or shape dissimilarity for passive haptics~\cite{de_tinguy2019.TangibleDissimilarity}.

\paragraph{Should all components have a coherent level of fidelity?}
The discrepancy in the levels of fidelity between different components can result in low experiential fidelity, as discussed above, or bring disadvantages in performance, preference, or other user experience aspects~\cite{bowman2012.Naturalism, mcmahan2016.UncannyInteration}. One could hypothesize that the lower fidelity is in one component of the \model, the more other components need to be enhanced to compensate. But it would also be reasonable to assume that other components must be matched at the same level to give the user a consistent impression. For example, in Mario Kart with cartoony visuals and a comical setting, high-fidelity vehicle control and physics would seem inappropriate, might decrease the perceived overall fidelity, and limit performance. Instead, the transfer functions and car behavior that make driving simple and error-tolerant lead to a coherent experience and arguably higher experiential fidelity. Similarly, we hypothesize that a driving simulator with sophisticated car physics and true-to-life input devices will be experienced as most realistic if the sensory feedback matches the high faithfulness and does not rely on cartoony visuals, funny sounds, or lacks haptics. 
Systematic experimental comparisons might reveal how uniform interactions should be designed.

\subsection{Methodology}

\paragraph{How can we quantify fidelity?}
Ideally, every component of the \model\ would come with means of quantification or a theoretically founded metric.
Due to the scope of this universal model and the depth of its conceivable subcomponents, achieving a set of methods for quantifying all fidelity components comprehensively is a considerable endeavor that the research community has worked on and will arguably need to continue working on for decades. 
\citet{al2022framework} proposed five categories in their framework for evaluating some fidelity aspects. The limitations of human sensory capabilities partially define the maximum. The other classifications are not delimited clearly. It is challenging to evaluate the moving target of rapidly evolving technology as it requires either dynamic adjustment of the classification or prospective universality. 
Ideally, a comprehensive framework would allow assigning numeric values objectively, reproducibly, and universally. Looking at haptic fidelity as an example, the Haptic Fidelity Framework by \citet{hapticfidelityframework} shows how complex and manifold it can be to specify even one of the subcomponents of display fidelity. This specialized framework identifies 14 factors defining haptic fidelity along the categories \textit{sensing}, \textit{hardware}, and \textit{software}. The publication includes an expert tool to quantify each factor and calculate an aggregated haptic fidelity score on a five-point Likert scale for technology-agnostic comparison. 
Considering that this covers only one modality in one out of eight fidelity components, it poses a significant research opportunity to provide such sophisticated frameworks for all aspects of interaction fidelity. Meanwhile, a validated questionnaire for users' self-reports in studies on haptic fidelity is still missing---leading to the following research opportunity.

\paragraph{How can we measure each aspect?}
Given we have means to describe all fidelity factors quantitatively, we need to establish methods, standards, and instruments to measure it objectively and subjectively. 
\citet{goncalves2021.ReviewRealismMethods} recently presented a systematic literature review on the methodology of 79 studies on VR realism. 
For heuristic expert analysis, specialized frameworks, such as the FIFA~\cite{mcmahan2016.UncannyInteration}, the Haptic Fidelity Framework\cite{hapticfidelityframework}, or the Simulation Fidelity Rating Scale for flight simulators~\cite{perfect2014.FlightSimulationFidelity} can help investigate single (sub)components, as outlined before. 
Furthermore, standards for technical evaluations are needed to compare devices, for instance, regarding the physical similarity of generated sensory stimuli or the accuracy of tracking devices. 
Validated instruments for psychometric evaluations allow comparison between studies on perceptual fidelity. 
Case-specific behavioral measures can improve our assessment of how realistically users react but are difficult to standardize. For subjective assessments, specialized and validated possibilities for self-reporting are essential beyond broad subscales of presence questionnaires. We propose developing dedicated tools to understand the perceived fidelity of the distinct interface components instead of generalizing overall realism.

\paragraph{Which specialized tools are out there?}
Gathering the already available resources on fidelity research would significantly support the community. However, comprehensively collecting and arranging all frameworks, instruments, questionnaires, etc., is challenging. This work provides a first step, particularly with the overview in Table~\ref{tab:signpost}. Beyond that, a systematic review is needed.

\paragraph{What should reporting guidelines include?}
Another crucial challenge for interaction fidelity research is finding standard reporting guidelines for fostering comparability and generalizability. Currently, relevant information about the system design is often missing to understand evaluation results and apply meta-analyses comprehensively. The \model\ might serve as a starting point to agree on reporting guidelines ensuring all system components that define interaction fidelity as a holistic concept are being reported. Because of the interdependencies of the single aspects, we encourage researchers to report details about the interactions beyond the element of interest.

\section{Conclusion}
To understand what makes interactions in VR simulations more or less faithful to the real world or any other reference frame, we must distinguish between various aspects of fidelity. In this article, we proposed the \fullmodel\ (\model) that allows analyzing how closely a virtual interaction corresponds to the original along various factors. We define eight fidelity components along the HCI loop: action, detection, transfer, simulation, rendering, display, perceptual, and experiential fidelity. 

The consequent terminology offered in this work supports precise communication and consistency across publications in the field. With a clear structure, rigorous explanations, practical examples, and a guideline with best practices, we demonstrate how VR professionals in research and development can use the conceptual model to describe, understand, compare, hypothesize, and teach. 
With its theoretically grounded simplicity, the \model\ can be universally applied to any VR experience. Therefore, our taxonomy defines only twelve general fidelity terms, as listed in Table~\ref{tab:definitions}. Beyond that, this article serves as a signpost referring readers to previous publications with specialized frameworks or concepts, as each component can be further distinguished in more detail.

The presented model underwent rigorous, critical discussions and was refined iteratively. For validation, we conducted 14 extensive interviews with experts from academia and the industry to review and test the model from different perspectives. The thematic analysis showed criticism of the concept, the terminology, and applications, identified application strategies, and outlined various benefits and use cases of the model. All experts found it interesting and helpful. 
As suggested in the interviews, we provide educational material as part of the supplemental material, including modern posters and a slide deck, which are free to use and adapt.

From our practical experiences with the model, we identified common patterns that might be prevalent in various use cases and interaction techniques. By connecting the dots in such a way, the \model\ will support finding similarities in study results and see their findings in a bigger picture. 
Using the model to think about the fidelity of VR interactions opens up promising opportunities for systematic and targeted research. We hope to inspire new directions in research for a better understanding of interactions in VR.

\begin{acks}
We are grateful to the experts for sharing their perspectives and criticism in the interviews. We thank Yvonne Rogers, Don Norman, and Katja Rogers for helping shape this work, and \href{https://www.canilvisuals.com}{Laura Canil} for realizing the visual designs of Figures~\ref{fig:loop} and~\ref{Fig:MVC}. 
\end{acks}

\subsection*{Funding}
This research was partially funded by the Klaus Tschira Stiftung.

\subsection*{Conflicts of Interest}
The authors report there are no competing interests to declare. 

\bibliographystyle{ACM-Reference-Format}
\bibliography{references}

\appendix

\section{Appendix: Educational Material}
\begin{figure*}[ht!]
  \centering
  \includegraphics[width=0.5\linewidth]{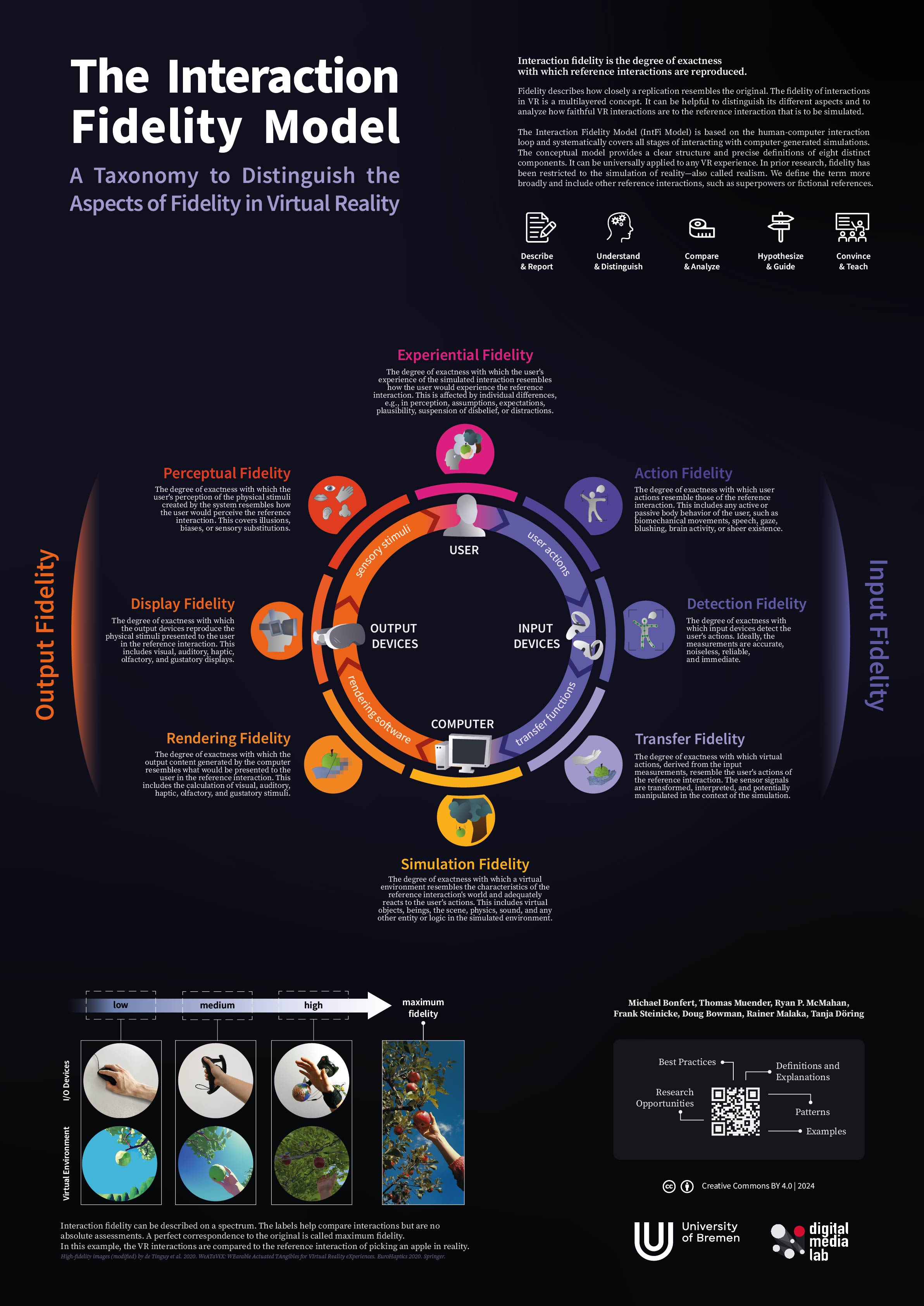}
  \caption{A preview of the large ISO A0 portrait poster. It will be distributed in the highest printing quality after peer reviewing.}
  \Description{A dark poster conveying the central concepts of this work.}
  \label{fig:appendix-poster-A0}
\end{figure*}

\begin{figure*}[ht!]
  \centering
  \includegraphics[width=0.5\linewidth]{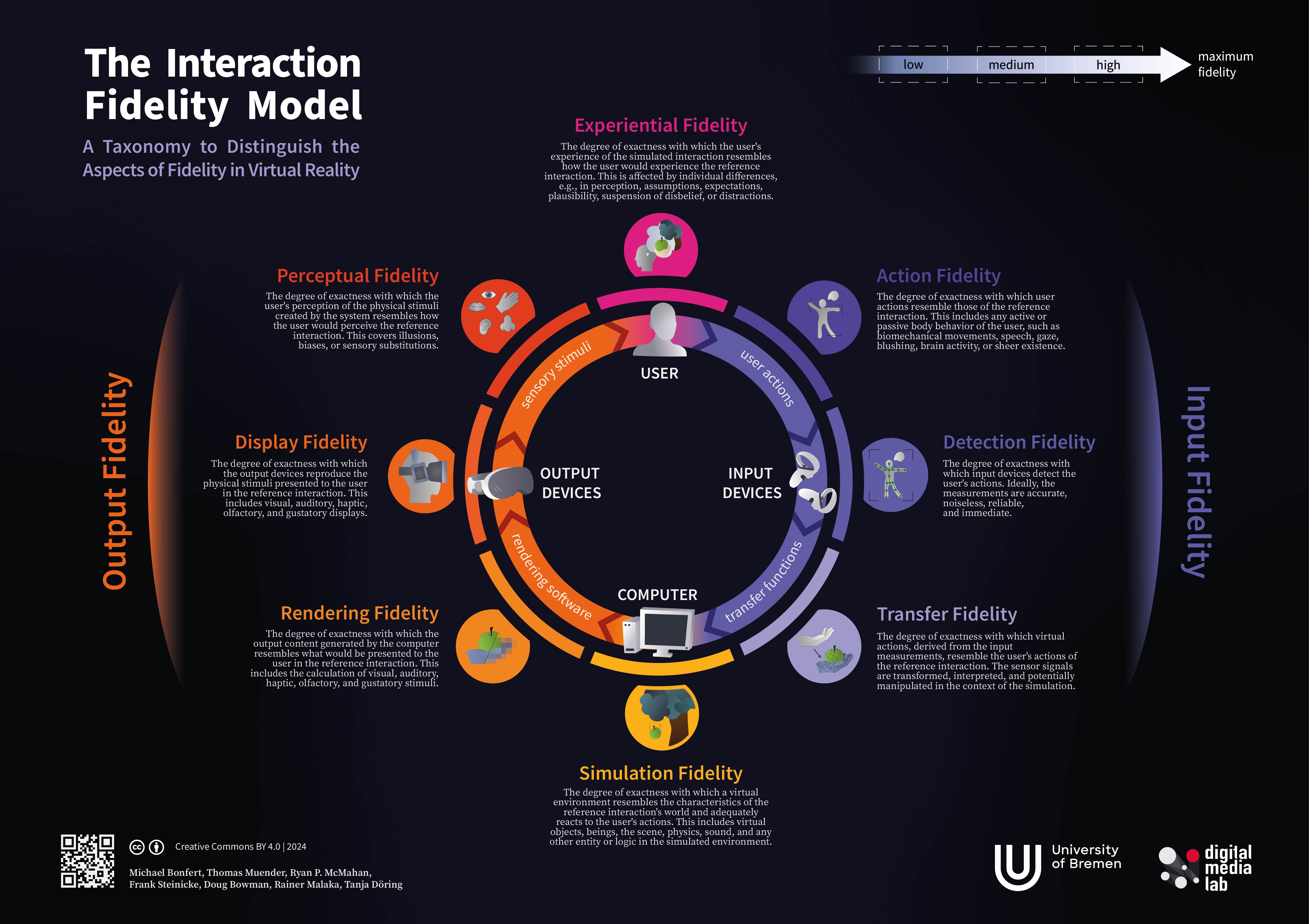}
  \caption{A preview of the ISO A1 landscape poster. It will be distributed in the highest printing quality after peer reviewing.}
  \Description{A dark poster with the central loop from \autoref{fig:loop}.}
  \label{fig:appendix-poster-A1}
\end{figure*}

\begin{figure*}[ht!]
  \centering
  \includegraphics[width=0.9\linewidth]{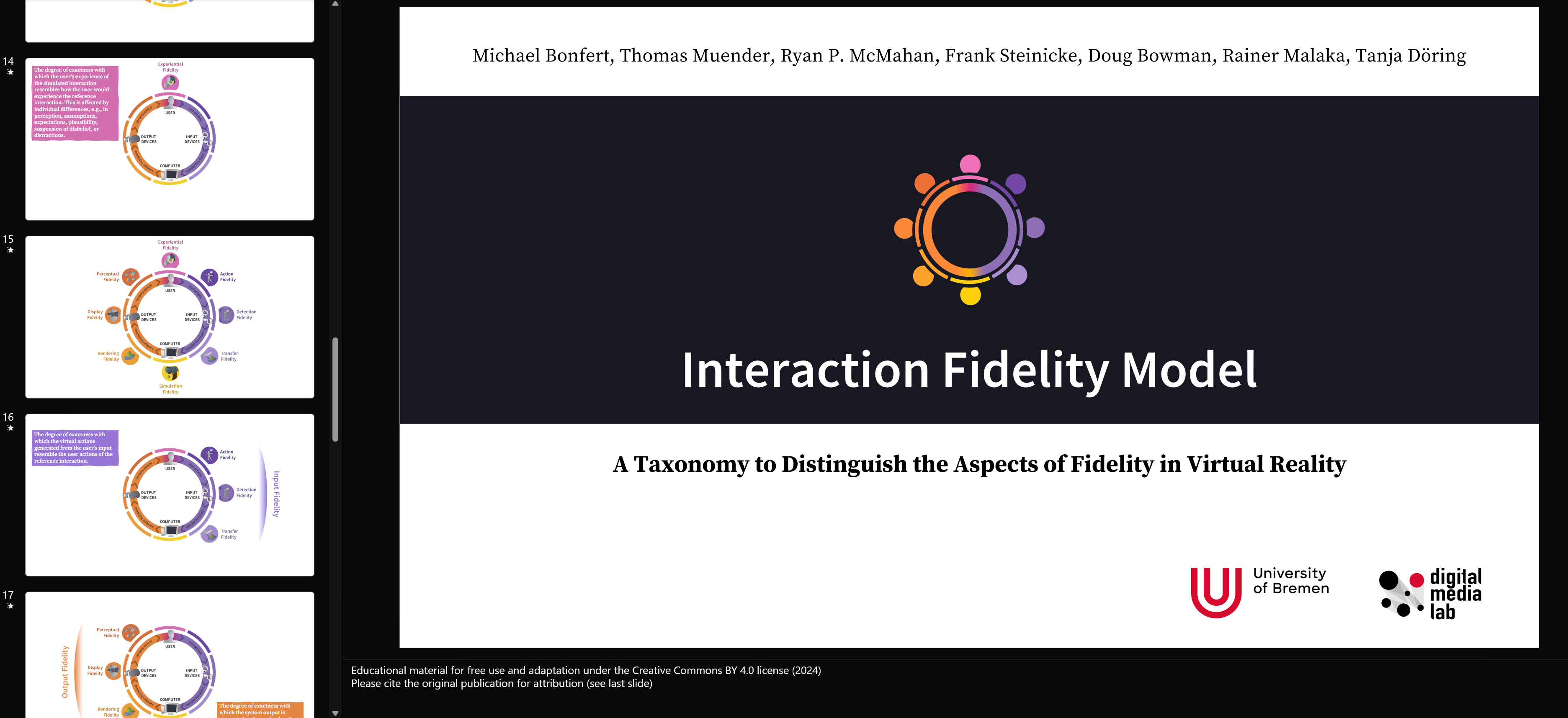}
  \caption{A preview of the slide deck outlining the central concepts of this work. The editable PPTX file will be distributed after peer reviewing.}
  \Description{The title slide and a preview of the upcoming slides on the side.}
  \label{fig:appendix-slides}
\end{figure*}

\begin{figure*}[ht!]
  \centering
  \includegraphics[width=1\linewidth]{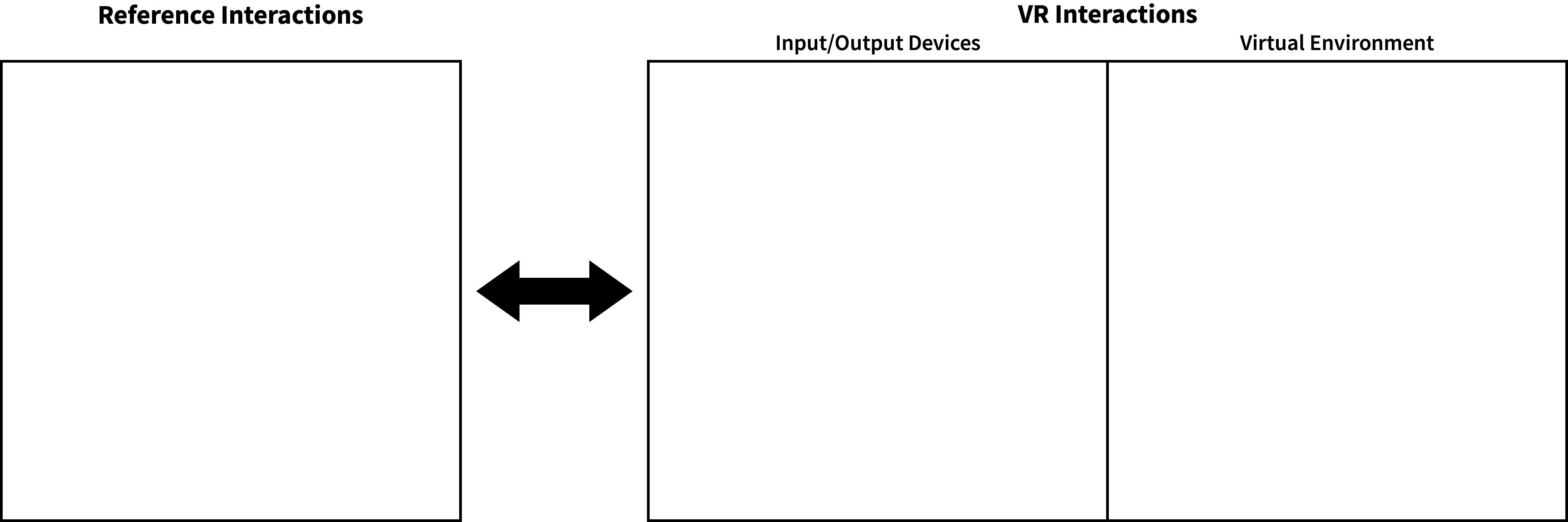}
  \caption{The template of the Correspondence Figure in which other systems can be inserted. The high-resolution PNG and an editable PSD file will be distributed after peer reviewing.}
  \Description{Boxes with space for the reference interaction, input/output devices, and the virtual environment with an arrow in between.}
  \label{fig:appendix-correspondence}
\end{figure*}

As part of this publication, we will provide educational material as supplemental material. It will be distributed under the Creative Commons BY 4.0 license (\url{https://creativecommons.org/licenses/by/4.0}). It can then be shared, adapted, and printed as long as the original publication and authors are appropriately cited, and any modifications are indicated. Figures~\ref{fig:appendix-poster-A0}, \ref{fig:appendix-poster-A1}, \ref{fig:appendix-slides}, and \ref{fig:appendix-correspondence} preview the files that will be made available upon acceptance of the peer-reviewed article.

We will further share a template of the Correspondence Figure as used in Figures~\ref{Fig:correspondence}, \ref{fig:lapsim}, \ref{fig:juggling}, and \ref{fig:lightsaber}, which compares a reference interaction to an implementation in VR. It can be adapted in other publications and will be available as PNG and PSD (Adobe Photoshop) files.

\end{document}